\documentclass{article}
\usepackage{cite}
\usepackage{a4wide}
\usepackage{graphicx}
\usepackage{wrapfig}
\usepackage{fancyhdr}
\usepackage{graphicx}
\usepackage[english]{babel}
\usepackage[bookmarks]{hyperref}
\usepackage{graphicx}
\usepackage{bm}
\usepackage[mathlines]{lineno}
\usepackage{subfigure} 
\usepackage{cancel}
\usepackage{amsmath} 
\usepackage{verbatim}
\usepackage{mathtools}
\usepackage{caption}
\usepackage[normalem]{ulem}
\usepackage{amssymb}
\usepackage{amstext} 
\usepackage{amsfonts} 
\usepackage{mathrsfs}
\usepackage{braket}
\usepackage{tikz}
\usepackage{wrapfig}
\usepackage{mwe}
\usepackage{float}

\DeclareMathOperator{\tr}{tr}

\usepackage[utf8]{inputenc}
\usepackage[left=2.5cm,right=2.5cm,top=2cm,bottom=2cm]{geometry}

\title{\vfill \textbf{The strong couplings of massive Yang-Mills theory}}
\author{\textbf{Anamaria Hell}\footnote{Hell.Anamaria@physik.uni-muenchen.de}}
\date{\textit{ Ludwig-Maximilians-Universität,\\
Theresienstraße 37, 80333 Munich, Germany}}
\begin{document}

\maketitle
\thispagestyle{empty} 

\begin{abstract}
We study the massive Yang-Mills theory in which the mass term is added \textit{by hand}. The standard perturbative approach suggests that the massless limit of this theory is not smooth. We confirm that this issue is related to the existence of additional degrees of freedom, which are absent in the massless theory. Nevertheless, we show that the longitudinal modes become strongly coupled at the Vainshtein scale, which coincides with the scale of the unitarity violation. Beyond this scale, they decouple from the remaining degrees of freedom, and the massless theory is restored up to small corrections. From here, it follows that the apparent discontinuity in the massless limit is only an artefact of the perturbation theory. The massless limit of massive Yang-Mills theory is smooth, as originally proposed in \cite{Vainshtein}.
\end{abstract}
\vspace*{\fill}

\clearpage
\pagenumbering{arabic} 
\newpage
\section{Introduction}
The massive Yang-Mills theory with mass added \textit{by hand} has served as one of the starting points in the construction of the Standard Model \cite{Glashow}, and was even considered as a possibility of avoiding the hierarchy problem \cite{Delburgo}. On first sight, it appears that this theory presents difficulties for small values of mass, when treated by the standard approaches. The propagator of the vector boson is singular in mass. Thus, at high energies, it approaches a constant, indicating a power-counting non-renormalizable theory. 
Nevertheless, such an argument might be misleading, as a similar propagator appears in the Proca theory \cite{Proca}, which has been shown to be renormalizable when coupled to a conserved source \cite{BoulGil}.
Therefore, as a first step in analysing renormalizability, most authors have applied the non-Abelian generalisation of a field redefinition \textit{a la Stueckelberg} \cite{Stueckelberg, Kunimasa}. This brings the propagator into a form indicating a power-counting renormalizable theory. It has been shown that while the theory is finite for one-loop diagrams, it is not renormalizable for two or more loops \cite{SalamKomar, UmezawaKamefuchi, Ionides, Salam, Boulware, Veltman1970}. Recently, it was argued that the theory could be renormalizable in the sense of an effective field theory \cite{Gegelia}. Nevertheless, most of the studies have indicated that besides renormalizability, another problem appears. At a first glance, the theory violates unitarity. In studying the unitarity of the massive Yang-Mills theory, which was initiated by the development of the cutting rules \cite{cuttingrules, Cutkosky}, there were two interesting cases, based on the degrees of freedom of the theory. In the first case, only the transverse modes were set on the external lines of the diagrams. This restriction was justified by the absence of additional degrees of freedom in the massless theory, the longitudinal modes. It has been shown that the theory is unitary up to a loop \cite{Veltman1968}. However, unitarity was violated at two loops \cite{VeltmanReiff}. In the second case, the longitudinal modes were set on the external lines of the diagrams. Then, the unitarity violation for energies $k_u\sim\frac{g}{m}$ was already evident from the tree diagrams \cite{Vainshtein, Bell, Cornwall, Smith}. Nevertheless, in \cite{Veltman1970}, it was found that for two or more loops, the singularities in the mass form a series in $\frac{g^2\Lambda^2}{m^2}$, where $\Lambda$ is the cut-off scale. The same series was also given in \cite{Veltman1968} while analysing the n-point functions which contained only the longitudinal modes. This raised a question as to whether the re-summation of these series could cure the mass singularities appearing within the perturbation theory. However, due to the following argument, this possibility was excluded in \cite{vDVZ}. Even when only the transverse modes are set on the external lines, and tree level and one-loop diagrams are considered, the massive theory does not smoothly approach the massless one for $m\to0$\cite{vDVZ, SlavFad, Wong}. The reason for this is the difference in the number of degrees of freedom in massive and massless theories, according to \cite{vDVZ}. 
From the perspective of physical continuity \cite{BassSch}, such a behaviour is unexpected. If we modify a certain theory through the introduction of a new parameter, once we take the limit back to the original theory, the observable effects should also smoothly approach those of the original theory. However, one might notice that a similar kind of issue arises in the context of massive gravity. In massive linearized gravity with the mass term of the Fierz-Pauli form \cite{FierszPauli}, the discontinuity appears already for tree level diagrams \cite{vDVZ,Zakharov, Iwasaki}. As a result, the predictions for Mercury's perihelion precession and the deflection of starlight differed from those obtained in General relativity. Similarly to massive Yang-Mills theory, the reason for this disagreement lies in the longitudinal mode. However, for massive gravity it was shown that this apparent discontinuity is resolved through the Vainshtein mechanism \cite{VainshteinMeh, PertNep}. The nonlinear interactions that were not considered in \cite{vDVZ,Zakharov, Iwasaki} cause the longitudinal mode to enter a strong coupling regime at the Vainshtein radius, where they become of the same order as the kinetic term. As it was demonstrated in \cite{Mimetic}, the essence of the Vainshtein scale lies in the minimal level of quantum fluctuations for the fields. Beyond it, the longitudinal modes decouple from the rest of matter and General relativity is restored up to small corrections \cite{Gruzinov}. In the case of massive Yang-Mills theory, a similar conjecture was made in \cite{Vainshtein} - \textit{"...it appears highly probable that outside perturbation theory, a continous zero-mass limit exists and the theory is renormalizable."}. Using a decomposition of the vector field which resulted in a Lagrangian containing non-polynomial terms, the authors of \cite{Vainshtein} have suggested that the massless limit is smooth, and that the theory might be renormalizable, if these non-polynomial terms are not perturbatively expanded. However, it was noted in \cite{Slavnov1972} that for such a conclusion to hold, it is not sufficient that all the terms which contain longitudinal modes vanish in the massless limit at the level of the Lagrangian, as there is no guarantee that the same applies to matrix elements. The subsequent studies were hence concerned with finding a way to treat the Lagrangian which contains non-polynomial terms. In a promising attempt, \cite{QuadriUnit,Ferrari, QuadriOneloop} have deduced conditions for unitarity in the Landau gauge theory and proposed a subtraction scheme for the divergences. However, these conditions cannot be applied to the unitary gauge as well \cite{Quad}. On the other hand, the authors of \cite{Dragon} have shown by applying several field redefinitions which do not alter the S-matrix, that the non-polynomial theory can be brought to the polynomial form resembling the one of \cite{Salam}. This implied that the theory is not renormalizable and unitary \cite{Ruegg}. However, if the non-polynomial theory can be algebraically reduced to a polynomial one, is that sufficient to conclude that the massless limit is not smooth? 
Even though a field redefinition does not necessarily change the S-matrix, it clearly changes the definition of the degrees of freedom. While the Lagrangian which is polynomial in fields corresponds to a linear decomposition of the vector field \cite{SalamKomar, UmezawaKamefuchi, Ionides, Salam}, the non-linear one introduces the non-polynomial terms in the Lagrangian \cite{Kunimasa}. 
Nevertheless, we will show that only the non-linear decomposition of the vector field properly defines the transverse and longitudinal modes. If we would have started with a linear one, we would encounter an infinite amount of strong coupling scales, which disagree with the scale of the unitarity violation, and also necessary field redefinitions.  
\\
The purpose of this paper is to show that the apparent violation of unitarity, and mass singularity of the perturbative series are just artefacts of the perturbation theory. First, we will show that these issues arise due to the properly defined longitudinal modes, which are absent in the massless theory. Within the perturbation theory, only these modes enter the strong coupling regime, at a Vainshtein scale which coincides with the scale of unitarity violation. Evaluating the theory beyond the strong coupling scale, we will find that the corrections to the transverse modes due to the longitudinal ones become suppressed by the strong coupling scale, and hence decrease as we approach higher energies. Therefore, we will show that the initial suggestion made in \cite{Vainshtein} was correct. The massless theory is recovered up to a small correction, and the massless limit is smooth.


\section{The basics of massive Yang-Mills theory}

In this paper we will study the simplest model of massive Yang-Mills theory, in which the mass term is added \textit{"by hand"} to the Yang-Mills theory\footnote{To address the problems of renormalizability and unitarity arising due to the propagator (\ref{eq::propagator}), \cite{FradkinTy} have initiated the consideration of various modified models of massive Yang-Mills theory (see e.g. \cite{Delburgo, Ruegg} for a review). However, these models will not be of interest in this paper. We will instead focus on the simplest possible model that is given by (\ref{eq::action}). }. The action of this model is given by 
\begin{equation}\label{eq::action}
    S=\int d^4x \left[-\frac{1}{2}\tr(F_{\mu\nu}F^{\mu\nu})+m^2\tr(A_{\mu}A^{\mu})\right],
\end{equation}
where 
\begin{equation}
      F_{\mu\nu}=D_{\mu}A_{\nu}-D_{\nu}A_{\mu}
\end{equation}
is the field strength tensor. The covariant derivative is given by  
\begin{equation}
    D_{\mu}=\partial_{\mu}+igA_{\mu}
\end{equation}
 and $g$ is the coupling constant. We will assume that $g\ll 1$. The vector field $A_{\mu}$ is the $2\times2$ hermitian matrix, which we will expand in terms of the generators of SU(2) as 
\begin{equation}
    A_{\mu}=A_{\mu}^aT^a,    \qquad \text{with}\qquad T^a=\frac{\sigma^a}{2} \qquad\text{and}\qquad a=1,2,3.
\end{equation}
Here, $\sigma^a$ are the Pauli matrices. Main cause for the discontinuity in the massless limit, along with the unitarity violation results from the propagator of the vector field. In momentum space it is given by 
\begin{equation}\label{eq::propagator}
    \tilde{\Delta}_{\mu\nu}^{ab}(k)=\left(-\eta_{\mu\nu}+\frac{k_{\mu}k_{\nu}}{m^2}\right)\frac{i\delta^{ab}}{k^2-m^2+i\epsilon}.
\end{equation}
The term containing the inverse power of mass is clearly problematic. At energy scales $k^2\gg m^2$ it tends to a constant and thus indicates a power counting non-renormalizable theory. In this chapter we will show that the origin of this term lies in the longitudinal mode. 
\\\\
We will start our analysis with the free theory, by setting the coupling constant to zero. In this case, the theory reduces to three Proca theories, one for each $a=1,2,3$. Our first goal is to analyse express the theory only in terms of the physical degrees of freedom. In order to do so, we will first separate the temporal and spatial components of the vector field. In addition, we will decompose the spatial part of the vector field according to the irreducible representations of SO(3) group:
\begin{equation}\label{eq::decomposition}
    A_i^a=A_i^{Ta}+\chi_{,i},\qquad\text{where}\qquad A_{i,i}^{Ta}=0.
\end{equation}
Here, commas denote the derivatives with respect to the corresponding coordinate $x^{\mu}$. The vectors $A_i^{Ta}$ are the transverse modes and scalars $\chi^a$ are the longitudinal modes. Substituting this decomposition into (\ref{eq::action}), we obtain 
\begin{equation}\label{eq::action2}
    \begin{split}
        S=\frac{1}{2}\int d^4x &\left[A_0^a\left(-\Delta+m^2\right)A_0^a+2A_0^a\Delta\dot{\chi}^a-\left(\dot{\chi}^a\Delta\dot{\chi}^a-m^2\chi^a\Delta\chi^a\right)\right.\\ &\left.+\left(\dot{A_i}^a\dot{A_i}^{Ta}-A_{i,j}^{Ta}A_{i,j}^{Ta}-m^2A_i^{Ta}A_i^{Ta}\right)\right].
        \end{split}
\end{equation}
The point in this equation denotes a time derivative. As none of the time derivatives act on the $A_0^a$ components, we can conclude that these are not propagating. Therefore, we will find the constraints that are fulfilled by them and insert the solutions back into the action. Then, we will obtain the action which consists only of the propagating degrees of freedom. By varying the action with respect to $A_0^a$, we find the following constraints
\begin{equation}
     (-\Delta+m^2)A_0^{a}=-\Delta\dot{\chi}^{a},
\end{equation}
whose solutions are given by
\begin{equation}\label{eq::const0}
    A^{a}_0=\frac{-\Delta}{-\Delta+m^2}\dot{\chi}^{a}.
\end{equation}
Here, the operator $\frac{1}{-\Delta+m^2}$ should be understood in terms of the Fourier modes. Substituting (\ref{eq::const0}) back into (\ref{eq::action2}), we obtain
\begin{equation}\label{eq::action3}
    S=-\frac{1}{2}\int d^4x\left[A_i^{Ta}(\Box+m^2)A_i^{Ta}+\chi^a(\Box+m^2)\frac{m^2(-\Delta)}{-\Delta+m^2}\chi^a\right].
\end{equation}
Therefore, we can see that this theory has three physical degrees of freedom per colour, two transverse modes and one longitudinal for each $a=1,2,3$. In this action the longitudinal mode is not canonically normalized. Defining the normalised variable as
\begin{equation}\label{eq::cannorm}
    \chi_n^a=\sqrt{\frac{-\Delta m^2}{-\Delta+m^2}}\chi^a,
\end{equation}
(\ref{eq::action3}) becomes 
\begin{equation}\label{eq::action4}
    S=-\frac{1}{2}\int d^4x\left[A_i^{Ta}(\Box+m^2)A_i^{Ta}+\chi_n^a(\Box+m^2)\chi_n^a\right].
\end{equation}
Let us now see from where the problematic term in the propagator (\ref{eq::propagator}) comes from. Since the modes are now canonically normalised, it is easy to read of the corresponding propagators from (\ref{eq::action4}). They are given by
\begin{equation}
    \begin{split}
        &\Delta_{\chi_n}^{ab}(k)=\frac{i\delta^{ab}}{k^2-m^2}\qquad\text{and}\qquad
        \Delta_{ij}^{Tab}(k)=\left(\delta_{ij}-\frac{k_ik_j}{|\vec{k}|^2}\right)\frac{i\delta^{ab}}{k^2-m^2},
    \end{split}
\end{equation}
where the first one corresponds to the propagator of the longitudinal modes, and the second to the propagator of the transverse modes. Using (\ref{eq::cannorm}), we can now obtain the propagator of the original longitudinal mode
\begin{equation}
    \Delta_{\chi}^{ab}(k)=\frac{|\vec{k}|^2+m^2}{m^2|\vec{k}|^2}\frac{i\delta^{ab}}{k^2-m^2}.
\end{equation}
This propagator has the same divergence in mass as (\ref{eq::propagator}). In fact, by considering separately the components of (\ref{eq::propagator}), we can easily see that these will reduce to the combinations of the propagators for the transverse and original longitudinal modes. Hence we can conclude that the longitudinal modes will be the cause of the discontinuity in the massless limit and the violation of unitarity. 
\\\\
In order to analyse the interacting theories in the following chapters, we will use the minimal level of the quantum fluctuations of the modes. In the case of a normalised quantum field, the minimal level of quantum fluctuations is given by $\left(\frac{k^3}{\omega_k}\right)^{\frac{1}{2}}$, where $\omega_k=\sqrt{k^2+m^2}$ is the frequency of each mode $\vec{k}$ \cite{QFTCS}. For energy scales $k^2\sim\frac{1}{L^2}\gg m^2$, where $L$ is the corresponding length-scale, the minimal level of quantum fluctuations for the transverse and normalized longitudinal modes is then given respectively by 
\begin{equation}\label{eq::quanfluct}
    \delta A_L^{Ta}\sim\frac{1}{L} \qquad\text{and}\qquad \delta\chi_{nL}^a\sim\frac{1}{L}, \quad\text{}\quad a=1,2,3.
\end{equation}
From here it follows that the minimal level of quantum fluctuations for the longitudinal mode is given by
\begin{equation}
    \delta\chi_L^a\sim\frac{1}{mL},\qquad a=1,2,3.
\end{equation}

\section{From linear to nonlinear decomposition}

We will now consider the interacting theory. We begin our analysis with the linear decomposition of the vector field (\ref{eq::decomposition}). As a first step, we will express the action (\ref{eq::action}) in terms of these modes. For this purpose, we will first resolve the constraints satisfied by the $A_0$ components and substitute them into the action. Varying the action (\ref{eq::action}) with respect to $A_0$, we obtain the following constraints: 
\begin{equation}\label{eq::constraint}
\begin{split}
    (-\Delta+m^2)A_0^a=-\dot{A}_{i,i}^a-g\epsilon^{abc}\dot{A}_i^bA_i^c-g\epsilon^{abc}A_i^bA_{0,i}^c-g\epsilon^{abc}\partial_i(A_i^bA_0^c)-g^2\epsilon^{fbc}\epsilon^{fad}A_0^bA_i^cA_i^d,
\end{split}
\end{equation}
whose solution can be evaluated up to $\mathcal{O}\left(g^2\right)$ as
\begin{equation}\label{eq::constraintsol}
\begin{split}
    A_0^a&=D\left[\dot{\chi}^a\right]
    -\frac{g\varepsilon^{abc}}{-\Delta+m^2}\left[\dot{A}_i^bA_i^c+(\Delta\chi^b+2A_i^b\partial_i)D\left[\dot{\chi}^c\right]\right]\\\\
    &+\frac{g^2\varepsilon^{fab}\varepsilon^{fcd}}{-\Delta+m^2}\left\{(\Delta\chi^b+2A_i^b\partial_i)\frac{1}{-\Delta+m^2}\left[\dot{A}_j^cA_j^d+(\Delta\chi^c+2A_j^c\partial_j)D\left[\dot{\chi}^d\right]\right]+A_i^bA_i^cD\left[\dot{\chi}^d\right]\right\}.
\end{split}
\end{equation}
Here, $\chi^a$ is the longitudinal component of $A_i^a$, that we decompose according to (\ref{eq::decomposition}), while the operator $D$ is given by 
\begin{equation}
    D[\chi]=\frac{-\Delta}{-\Delta+m^2}\chi.
\end{equation}
Substituting (\ref{eq::constraintsol}) into the action (\ref{eq::action}), we obtain the Lagrangian density 
\begin{equation}\label{eq::Llin}
    \mathcal{L}=\mathcal{L}_0+\mathcal{L}_{int},\qquad\text{where}
\end{equation}
\begin{equation*}
    \begin{split}
        \mathcal{L}_0&=-\frac{1}{2}\chi^a(\Box+m^2)\frac{m^2(-\Delta)}{-\Delta+m^2}\chi^a-\frac{1}{2}A_i^{Ta}(\Box+m^2)A_i^{Ta}\qquad\text{and}\\\\
        \mathcal{L}_{int}&=g\varepsilon^{abc}\left[\frac{1}{2}\chi^b\chi^c_{,i}\Box A_i^{Ta}-\dot{\chi}^a\chi^{c}_{,i}\frac{m^2}{\Delta}\left(\dot{\chi}^b_{,i}\right)+\chi^bA_i^{Tc}\Box A_i^{Ta}\right]\\\\
        &+\frac{g^2}{2}\varepsilon^{fab}\varepsilon^{fcd}\left[\dot{\chi}^a\dot{\chi}^c\chi^b_{,i}\chi^{d}_{,i}+\left(\dot{\chi}^a_{,i}\chi^b_{,i}+\dot{\chi}^a\Delta\chi^b\right)\frac{1}{\Delta}\left(\dot{\chi}^c_{,j}\chi^d_{,j}+\dot{\chi}^c\Delta\chi^d\right)-\frac{1}{2}\chi^a_{,i}\chi^c_{,i}\chi^b_{,j}\chi^d_{,j}\right]\\\\
        &+\mathcal{O}\left(\frac{gA^{T3}}{L},\frac{gm^2A^T\chi^2}{L^3}, \frac{g^2A^T\chi^3}{L^3}\right),
    \end{split}
\end{equation*}
evaluated on scales $\frac{1}{L^2}\sim k^2\gg m^2$. We have retained only the most significant terms and denoted derivatives in the last line with $\frac{1}{L}$. The complete expression up to and including $\mathcal{O}\left(g^2\right)$ can be found in the Appendix. As in the free theory, the kinetic term for the longitudinal modes is not canonically normalised. With respect to the normalised variable defined in (\ref{eq::cannorm}), the Lagrangian density is given by
\begin{equation}\label{eq::Lnormlin}
    \mathcal{L}=\mathcal{L}_0+\mathcal{L}_{int},\qquad\text{where}
\end{equation}
\begin{equation*}
    \begin{split}
        \mathcal{L}_0&=-\frac{1}{2}\chi^a_n(\Box+m^2)\chi^a_n-\frac{1}{2}A_i^{Ta}(\Box+m^2)A_i^{Ta}\qquad\text{and}\\\\
        \mathcal{L}_{int}&\sim g\varepsilon^{abc}\left[\frac{1}{2m^2}\chi_n^b\chi^c_{n,i}\Box A_i^{Ta}-\frac{1}{m}\dot{\chi}_n^a\chi^{c}_{n,i}\frac{1}{\Delta}\left(\dot{\chi}^b_{n,i}\right)+\frac{1}{m}\chi_n^bA_i^{Tc}\Box A_i^{Ta}\right]\\\\
        &+\frac{g^2}{2m^4}\varepsilon^{fab}\varepsilon^{fcd}\left[\dot{\chi}_n^a\dot{\chi}_n^c\chi^b_{n,i}\chi^{d}_{n,i}+\left(\dot{\chi}^a_{n,i}\chi^b_{n,i}+\dot{\chi}_n^a\Delta\chi^b_n\right)\frac{1}{\Delta}\left(\dot{\chi}^c_{n,j}\chi^d_{n,j}+\dot{\chi}^c_n\Delta\chi^d_n\right)-\frac{1}{2}\chi^a_{n,i}\chi^c_{n,i}\chi^b_{n,j}\chi^d_{n,j}\right].
    \end{split}
\end{equation*}
Inverse powers of mass indicate that at high energies unitarity will be violated. According to \cite{Vainshtein}, this corresponds to length scales $L_u\sim\frac{g}{m}$. Let us now determine the strong coupling scale of the theory. It is the scale for which the non-linear terms are of the same order as the linear terms with respect to the equation of motion. Varying (\ref{eq::Llin}) with respect to $\chi$, we find the equation of motion satisfied by the longitudinal modes to be\footnote{The results are independent of whether the original or the normalised longitudinal modes are used.}
\begin{equation}\label{eq::Longlineom}
    \begin{split}
        (\Box+m^2)\chi^a&\sim\frac{g}{m^2}\varepsilon^{abc}\left[\chi^b_{,i}\Box A_i^{Tc}+A_i^{Tb}\Box A_i^{Tc}+m^2\left(\chi^c_{,i}\frac{1}{\Delta}\Ddot{\chi}^b_{,i}+\Ddot{\chi}^b\chi^c_{,i}+\Ddot{\chi}^b\Delta\chi^c+\dot{\chi}^b\Delta\dot{\chi}^c\right)\right]\\\\
        &+\frac{g^2}{m^2}\varepsilon^{fab}\varepsilon^{fcd}\chi^b_{,i}\left[\frac{\partial_i\partial_0}{\Delta}\left(\dot{\chi}^c_{,j}\chi^d_{,j}+\dot{\chi}^c\Delta\chi^d\right)-\dot{\chi}^c\dot{\chi}^d_{,i}-\Ddot{\chi}^c\chi^d_{,i}+\chi^d_{,i}\Delta\chi^c+\chi^c_{,j}\chi^d_{,ij}\right].
    \end{split}
\end{equation}
Varying (\ref{eq::Llin}) with respect to the transverse modes, we obtain 
\begin{equation}\label{eq::Translineom}
    (\Box+m^2)A_i^{Ta}\sim g\varepsilon^{abc}P_{ij}^T\left[\frac{1}{2}\Box\left(\chi^b\chi^c_{,j}\right)+A_j^{Tc}\Box\chi^b+2\chi^b_{,\mu}A_j^{Tc,\mu}\right],
\end{equation}
where 
\begin{equation}
    P_{ij}^T=\delta_{ij}-\frac{\partial_i\partial_j}{\Delta}
\end{equation}
is the transverse projector. In both (\ref{eq::Longlineom}) and (\ref{eq::Translineom}) we have retained only the relevant terms. We will now develop the perturbation theory by first expanding the fields into powers of the coupling constant: 
\begin{equation}
    A_i^T=A_i^{T(0)}+A_i^{T(1)}+...\qquad\text{and}\qquad \chi=\chi^{(0)}+\chi^{(1)}+...
\end{equation}
The modes $A_i^{T(0)}$ and $\chi^{(0)}$ satisfy free equations of motion
\begin{equation}
    (\Box+m^2)A_i^{Ta(0)}=0\qquad\text{and}\qquad (\Box+m^2)\chi^{a(0)}=0,
\end{equation}
whose solutions are plane waves. We will first analyse the transverse modes. The first-order corrections are given by 
\begin{equation}\label{eq::TransCor}
    (\Box+m^2)A_i^{Ta(1)}\sim g\varepsilon^{abc}P_{ij}^T\left(\chi^{b(0)}_{,\mu}\chi^{c(0),\mu}_{,j}+2\chi^{b(0),\mu}A_{i,\mu}^{Tc(0)}\right).
\end{equation}
 Estimating $\partial_{\mu}\sim\frac{1}{L}$, we can evaluate these terms as 
\begin{equation}
   g\varepsilon^{abc}P_{ij}^T\left( \chi^{b(0)}_{,\mu}\chi^{c(0),\mu}_{,j}\right)\sim\frac{g\chi^2}{L^3}\qquad\text{and}\qquad  g\varepsilon^{abc}P_{ij}^T\left(\chi^{b(0),\mu}A_{i,\mu}^{Tc(0)}\right)\sim\frac{g\chi A^T}{L^2}.
\end{equation}
Taking into account the minimal level of quantum fluctuations for the modes (\ref{eq::quanfluct}), we can estimate these terms further as
\begin{equation}
    \frac{g\chi^2}{L^3}\sim\frac{g}{\left(mL\right)^2L^3}\qquad\text{and}\qquad\frac{g\chi A^T}{L^2}\sim\frac{g}{\left(mL\right)L^3}.
\end{equation}
Since we consider scales $\frac{1}{L^2}\gg m^2$, the first term among these is most dominant. Therefore, we can evaluate the first-order corrections for the transverse modes as  
\begin{equation}
    A_i^{Ta(1)}\sim\frac{g}{\left(mL\right)^2L}.
\end{equation}
The scale of the strong coupling is the scale at which these corrections become of the same order as the linear term $A^{T(0)}\sim\frac{1}{L}$. Comparing the two terms, we find that the transverse modes enter the strong coupling regime at scales  
\begin{equation}
    L_{str}^T\sim\frac{\sqrt{g}}{m}.
\end{equation}
This scale is larger than the scale characterising the breakdown of unitarity, $L_u\sim\frac{g}{m}$, which indicates that the transverse modes enter the strong coupling regime before the unitarity is violated. Even though these scales usually coincide, their mismatch should not be surprising given that they are physically distinct. While the scale of strong coupling arises at the level of the equations of motion, the scale of unitarity violation is defined by the scattering amplitudes. Still, one might suspect that the results in terms of transverse and longitudinal modes are not necessarily in agreement with the earlier work of \cite{vDVZ, Veltman1968, Veltman1970}. Nonetheless, in the Appendix, we will reproduce the discontinuity at one loop for the transverse modes and show that the scale $L_{str}^T$ does not appear for either one-loop or two-loop corrections to the propagator of the transverse modes. 
The most dominant contribution to the first-order corrections for the longitudinal modes is given by 
\begin{equation}
    \begin{split}
        (\Box+m^2)\chi^{a(1)}\sim g\varepsilon^{abc}\chi^{b(0)}_{,\mu}\Delta\chi^{c(0),\mu},
    \end{split}
\end{equation}
and can be estimated as
\begin{equation}\label{eq::cor}
    \chi^{a(1)}\sim\frac{g}{(mL)^2}.
\end{equation}
When these corrections become of the same order as $\chi^{(0)}\sim\frac{1}{mL}$, the longitudinal mode enters the strong coupling regime. This corresponds to the scale
\begin{equation}
    L_{str}\sim\frac{g}{m},
\end{equation}
which is smaller than $L_{str}^T$. Since the transverse modes enter the strong coupling regime ahead of the longitudinal modes, the corrections (\ref{eq::cor}) are no longer trustworthy. Nevertheless, this should be taken with more caution, since there could be terms from higher order corrections that might cause the strong coupling of longitudinal modes on the same scale $L_{str}^T$. For second-order corrections, this possibility arises from the following terms:
\begin{equation}\label{eq::longCor}
    (\Box+m^2)\chi^{a(2)}\sim \frac{g}{m^2}\varepsilon^{abc}\chi^{b(0)}_{,i}\Box A_i^{Tc(1)}-\frac{g^2}{m^2}\varepsilon^{fab}\varepsilon^{fcd}P^T_{ij}\left(\chi^{d(0),\mu}_{,j}\chi^{c(0)}_{,\mu}\right)
\end{equation}
First, we will set the transverse modes to zero. In this case, only the second term remains and we can estimate these corrections as follows:
\begin{equation}
    \chi^{(2)}\sim\frac{g^2}{\left(mL\right)^5}
\end{equation}
On the scale $L^T_{str}$, they are of the same order as the linear term, and as a result, the longitudinal mode enters the strong coupling regime. 
Below this scale, the minimal level of quantum fluctuations of the longitudinal mode is determined by the $\mathcal{O}\left(g^2\right)$ terms in the Lagrangian. We can estimate these terms as 
\begin{equation}
\mathcal{L}_{int}\supset\frac{g^2\chi^4}{L^4}.
\end{equation}
If we define the new canonically normalised variable $\chi_n\sim \frac{g}{L}\chi^2$, we find that the minimal level of quantum fluctuations for the longitudinal mode in the strong coupling regime is given by 
\begin{equation}\label{eq::Qfluct1}
    \delta\chi^a_L\sim\frac{1}{\sqrt{g}}.
\end{equation}
Let us now consider the case with transverse modes. Substituting (\ref{eq::TransCor}) into (\ref{eq::longCor}) cancels the term which caused the strong coupling of the longitudinal modes at scales $L^T_{str}$. This implies that the longitudinal modes enter the strong coupling regime at scales lower than $L^T_{str}$. However, as soon as the transverse modes enter the strong coupling regime, the relation (\ref{eq::TransCor}) is no longer valid. Thus, the previous cancellation does not take place and the second-order contributions to the longitudinal modes at the strong coupling scale $L_{str}^T$ are given by 
\begin{equation}
    \chi^{(2)}\sim\frac{g^2}{\left(mL_{str}^T\right)^5}.
\end{equation}
This is of the same order of magnitude as the linear term for the longitudinal modes. Therefore, the longitudinal modes also enter the strong coupling regime at the same scale as the transverse modes. Below it, the most dominant terms in the interacting part of the Lagrangian density are given by 
\begin{equation}
    \begin{split}
        \mathcal{L}_{int}&\sim\frac{g}{2}\varepsilon^{abc}\chi^b\chi^c_{,i}\Box A_i^{Ta}
        +\frac{g^2}{2}\frac{\chi^4}{L^4}.
    \end{split}
\end{equation}
From the last term, it follows that the minimum level of quantum fluctuations for scales $L\leq L_{str}^T$ is given by (\ref{eq::Qfluct1}). The quantum fluctuations for the transverse modes are then determined by the first term. Since  we can now evaluate it as
\begin{equation}
    \frac{g}{2}\varepsilon^{abc}\chi^b\chi^c_{,i}\Box A_i^{Ta}\sim\frac{1}{L^3}A^T,
\end{equation}
we see that the minimum level of quantum fluctuations remains the same as before. However, it can be seen that this term has the same order of magnitude as the kinetic term for the transverse modes, whereas in the case of the longitudinal modes, the latter term is more dominant than the kinetic term. As a consequence, the transverse modes for scales $L\leq L_{str}^T$ are no longer properly defined. We can also see this from the equation of motion (\ref{eq::Translineom}). The first term on the left-hand side is the most dominant, and beyond the strong coupling scale it can be evaluated as follows 

\begin{equation}
   \frac{g}{2}\varepsilon^{abc}P_{ij}^T\Box\left(\chi^b\chi^c_{,j}\right)\sim\frac{1}{L^3}.
\end{equation}
As this term has the same order as the linear one for the transverse modes at scales below $L^T_{str}$, we should include it in the kinetic term. In other words, we should redefine the transverse modes as  
\begin{equation}\label{eq::redef}
    A_i^{Ta}=B_i^{Ta}+\frac{g}{2}\varepsilon^{abc}P_{ij}^T\left(\chi^b\chi_{,j}^c\right),\qquad B_{i,i}^{Ta}=0,
\end{equation}
and repeat the whole procedure for the new transverse modes $B_i^{Ta}$. As soon as we replace the old transverse modes by the new ones, the terms that induced a strong coupling scale $L^T_{str}$ will drop out of the Lagrangian. However, a new strong coupling scale will then appear,
\begin{equation}
    \tilde{L}^T_{str}\sim\frac{g^{2/3}}{m},
\end{equation} 
which will require another redefinition. This shows us that even if we start with a linear decomposition of the vector field in  transverse and longitudinal modes, the theory leads to a non-linear decomposition.
\\\\
The analysis so far has shown us that the fields are not correctly defined if the decomposition of the vector field is linear in fields. This is due to the fact that both transverse and longitudinal modes enter the strong coupling regime at a scale which does not match with that of the unitarity violation, as shown in the Appendix. Although it is unphysical for the transverse modes to become strongly coupled, this is resolved for scales $L<L_{str}^T$, since the nonlinear term causing the strong coupling remains of the same order as the linear term. In other words, this means that the transverse modes have to be redefined. In fact, the redefinition will be required until the strong coupling scale coincides with the scale of unitarity violation. Thus, even if we start with a linear decomposition, we will end up with a non-linear one. In the next chapter, we will see that the final decomposition introduces non-polynomial terms into the Lagrangian. 
This clarifies the conclusions that have emerged from the analysis of \cite{Dragon}. The fact that the Lagrangian resulting from a nonlinear decomposition can be algebraically transformed into polynomial form is not sufficient to conclude that the theory violates unitarity. The polynomial form corresponds to fields which are not correctly defined.

\section{The final strong coupling scale and beyond}
Now we will consider the full non-linear decomposition of the vector field into transverse and longitudinal modes. We will decompose the spatial part of the vector field as 
\begin{equation}\label{eq::Nonlindec}
    A_i=\zeta A_i^T\zeta^{\dagger}+\frac{i}{g}\zeta_{,i}\zeta^{\dagger}.
\end{equation}
Here, $\zeta$ is the unitary matrix given by
\begin{equation}
    \zeta=e^{-ig\chi},
\end{equation}
and the transverse modes satisfy
\begin{equation}
    A_{i,i}^{T}=0.
\end{equation}
Expanding $\zeta$, the spatial component of the vector field becomes \begin{equation}\label{eq::Vainsh}
    \begin{split}
        A_i^a=A_i^{Ta}+\chi_{,i}^a-g\varepsilon^{abc}(A_i^{Tb}\chi^c+\frac{1}{2}\chi_{,i}^b\chi^c)-\frac{g^2}{2}\varepsilon^{fab}\varepsilon^{fcd}(\chi^bA_i^{Tc}\chi^d+\frac{1}{3}\chi^b\chi^c_{,i}\chi^d).
    \end{split}
\end{equation}
We can see that the first redefinition of the transverse modes (\ref{eq::redef}) coincides with the transverse part of the second term at $\mathcal{O}\left(g\right)$. In contrast, both the longitudinal and transverse modes are now properly defined. If we denote the fields of the linear decomposition from the previous section as $\tilde{\chi}$ and $\tilde{A}_i^T$, the relation between the two pairs of fields is as follows: 
\begin{equation}
    \tilde{\chi}=\frac{\partial_i}{\Delta}\left[\zeta A_i^T\zeta^{\dagger}+\frac{i}{g}\zeta_{,i}\zeta^{\dagger}\right]\qquad\text{and}\qquad \tilde{A}_i^T=P_{ij}^T\left[\zeta A_j^T\zeta^{\dagger}+\frac{i}{g}\zeta_{,j}\zeta^{\dagger}\right].
\end{equation}
Following the same procedure as before, we will now express the action (\ref{eq::action}) only in terms of the physical degrees of freedom. 
In matrix notation, the constraint (\ref{eq::constraint}) satisfied by the $A_0$ component has the following form: 
\begin{equation}\label{eq::constraint1}
    \begin{split}
        \left(-\Delta+m^2\right)A_0&=-\dot{A}_{i,i}+ig[\dot{A}_i,A_i]
        +ig\left(2\left[A_i,A_{0,i}\right]+\left[A_{i,i},A_0\right]\right)+g^2\left[A_i,\left[A_0,A_i\right]\right],
    \end{split}
\end{equation}
where $\left[\,\,,\,\right]$ is the commutator. The solution of (\ref{eq::constraint1}) is given by 
\begin{equation}\label{eq::Csolution}
    A_0=\zeta\frac{1}{D}\left(-\frac{i}{g}m^2\zeta^{\dagger}\dot{\zeta}+ig\left[\dot{A}_i^T,A_i^T\right]\right)\zeta^{\dagger}+\frac{i}{g}\dot{\zeta}\zeta^{\dagger}.
\end{equation}
with
\begin{equation}
    \frac{1}{D}=\frac{1}{-\Delta+m^2-2ig[A_i^T,\partial_i\bullet\;]+g^2[A_i^T,[A_i^T,\bullet\;]]}.
\end{equation}
Here, $\bullet$ denotes the place where the expression acted upon by $\frac{1}{D}$ is to be inserted, once $\frac{1}{D}$ is perturbatively evaluated. Substituting (\ref{eq::Csolution}) into the action (\ref{eq::action}), we obtain the Lagrangian density
\begin{equation}\label{eq::nonpertLagrangian}
    \mathcal{L}=\mathcal{L}^T_0+\mathcal{L}_0^{\chi}+\mathcal{L}^T_{int}+\mathcal{L}^{T\chi}_{int}
\end{equation}
\begin{equation*}
    \begin{split}
        &\mathcal{L}^T_0=\tr\left(\dot{A}_i^T\dot{A}_i^T-A_{i,j}^TA_{i,j}^T-m^2A_i^TA_i^T\right)\\\\
        &\mathcal{L}_0^{\chi}=-\frac{m^2}{g^2}\tr\left[\zeta^{\dagger}\dot{\zeta}\frac{-\Delta}{-\Delta+m^2}\left(\zeta^{\dagger}\dot{\zeta}\right)-\zeta^{\dagger}\zeta_{,i}\zeta^{\dagger}\zeta_{,i}\right]\\\\
        & \mathcal{L}^{T\chi}_{int}=\frac{2im^2}{g}\tr\left\{-A_i^T\zeta^{\dagger}\zeta_{,i}+m^2\zeta^{\dagger}\dot{\zeta}\frac{1}{D}\left[A_i^T,\frac{1}{-\Delta+m^2}\partial_i(\zeta^{\dagger}\dot{\zeta})\right]\right\}\\\\
        &-m^2\tr\left\{\zeta^{\dagger}\dot{\zeta}\frac{1}{D}[\dot{A}_i^T,A_i^T]+[\dot{A}_i^T,A_i^T]\frac{1}{D}(\zeta^{\dagger}\dot{\zeta})+m^2\zeta^{\dagger}\dot{\zeta}\frac{1}{D}\left[A_i^T,\left[A_i^T,\frac{1}{-\Delta+m^2}(\zeta^{\dagger}\dot{\zeta})\right]\right]\right\}\\\\
        &\mathcal{L}^T_{int}=\tr\left\{-2igA_i^TA_j^T\left(A_{j,i}^T-A_{i,j}^T\right)+g^2\left[\dot{A}_i^T,A_i^T\right]\frac{1}{D}\left[\dot{A}_j^T,A_j^T\right]+g^2\left(A_i^TA_j^TA_i^TA_j^T-A_i^TA_i^TA_j^TA_j^T\right)\right\}.
    \end{split}
\end{equation*}
We can see that now all terms containing the longitudinal modes appear multiplied by a mass term. If we set $m=0$, we obtain the massless theory, which agrees with \cite{Vainshtein}. 
\subsection{The strong coupling scale}
In order to verify the Vainshtein scale within the perturbative regime we will first expand $\frac{1}{D}$ and $\zeta$. From the kinetic term of transverse modes, we can easily see that the expansion in $\frac{1}{D}$ is possible at high energies as long as $g\ll 1$. On the other hand, the expansion in $\zeta$ stops being valid for $\chi\sim\frac{1}{g}$. Assuming that $\chi<\frac{1}{g}$, and expanding both $\frac{1}{D}$ and $\zeta$, we obtain the Lagrangian density at energies $k^2\gg m^2$
\begin{equation}\label{eq::Ldenspert}
    \mathcal{L}=\mathcal{L}_{0}+\mathcal{L}_{int}, \qquad\text{where}
\end{equation}
\begin{equation*}
    \begin{split}
        \mathcal{L}_0&=\tr\left[-\chi\left(\Box+m^2\right)\frac{-\Delta m^2}{-\Delta+m^2}\chi-A_i^T\left(\Box+m^2\right)A_i^T\right]\\\\
        \mathcal{L}_{int}&\sim\tr\left\{igm^4[\chi,\dot{\chi}]\frac{1}{\Delta}(\dot{\chi})-2igm^2A_i^T\chi\chi_{,i}-2ig\left[\dot{A}_i^T,A_i^T\right]\frac{m^2}{\Delta}\left(\dot{\chi}\right)-2igA_i^TA_j^T(A_{j,i}^T-A_{i,j}^T)\right.\\\\
        &\left.+\frac{m^2g^2}{6}\left(\chi_{,\mu}\chi\chi^{,\mu}\chi-\chi_{,\mu}\chi^{,\mu}\chi^2\right)-\frac{g^2m^2}{3}A_i^T\left[\chi,\left[\chi_{,i},\chi\right]\right]\right\}.
    \end{split}
\end{equation*}
Here, we have kept only the most relevant terms. For simplicity, we will analyse the terms of the Lagrangian with the original longitudinal modes, which are connected with the canonically normalised ones via (\ref{eq::cannorm}). Estimating the derivatives as $\partial_{\mu}\sim\frac{1}{L}$, we can evaluate the interacting terms as
\begin{equation}
    \begin{split}
        &igm^4[\chi,\dot{\chi}]\frac{1}{\Delta}(\dot{\chi})\sim gm^4\chi^3,\qquad 2igm^2A_i^T\chi\chi_{,i}\sim\frac{gm^2A^T\chi^2}{L},\qquad2ig\left[\dot{A}_i^T,A_i^T\right]\frac{m^2}{\Delta}\left(\dot{\chi}\right)\sim gm^2A^{T2}\chi \\\\
        &2igA_i^TA_j^TA_{j,i}^T\sim\frac{gA^{T3}}{L},\qquad \frac{m^2g^2}{6}\chi_{,\mu}\chi\chi^{,\mu}\chi\sim\frac{g^2m^2\chi^4}{L^2}\quad\text{and }\quad\frac{g^2m^2}{3}A_i^T\left[\chi,\left[\chi_{,i},\chi\right]\right]\sim\frac{g^2m^2A^T\chi^3}{L}.
    \end{split}
\end{equation}
Taking into account the minimal level of quantum fluctuations for the fields (\ref{eq::quanfluct}), these terms become
\begin{equation}
    \begin{split}
        &gm^4\chi^3\sim\frac{gm}{L^3},\qquad \frac{gm^2A^T\chi^2}{L}\sim\frac{g}{L^4},\qquad gm^2A^{T2}\chi\sim\frac{gm}{L^3} \\\\
        &\frac{gA^{T3}}{L}\sim\frac{g}{L^4},\qquad \frac{g^2m^2\chi^4}{L^2}\sim\frac{g^2}{\left(mL\right)^2L^4}\quad\text{and }\quad\frac{g^2m^2A^T\chi^3}{L}\sim\frac{g^2}{\left(mL\right)L^4}.
    \end{split}
\end{equation}
The terms at $\mathcal{O}(g)$ will always be smaller than the kinetic terms for the longitudinal and transverse modes due to the assumption for the coupling constant $g\ll1$, and since we are considering scales $\frac{1}{L^2}\sim m^2$. However, this does not hold for the terms at $\mathcal{O}(g^2)$. Among these, the most dominant term contains the quartic power of the longitudinal mode. If one compares it with the kinetic term, one then finds that the longitudinal modes enter the strong coupling regime at scales
\begin{equation}
    L_{str}\sim\frac{g}{m}.
\end{equation}
For the transverse modes, the last term is most dominant. This term becomes of the same order as the kinetic term at scale
\begin{equation}
    L\sim\frac{g^2}{m},
\end{equation}
which is smaller than $L_{str}$, and is therefore not trustworthy. Given that the most dominant nonlinear terms correspond to $\mathcal{O}\left(g^2\right)$, we will confirm the strong coupling scale also on the level of the equations of motion. Varying the action corresponding to the Lagrangian density (\ref{eq::Ldenspert}) with respect to the longitudinal modes, we obtain
\begin{equation}
    \begin{split}
        \left(\Box+m^2\right)\chi&\sim 2ig\left[A_i^T,\chi_{,i}\right]-ig\frac{m^2}{\Delta}\left(\left[\chi,\Ddot{\chi}\right]\right) +igm^2\left\{2\left[\dot{\chi},\frac{1}{\Delta}\left(\dot{\chi}\right)\right]+\left[\chi,\frac{1}{\Delta}\left(\Ddot{\chi}\right)\right]\right\}\\\\
        &+2ig\frac{1}{\Delta}\left(\left[\Ddot{A}_i^T,A_i^T\right]\right)+\frac{g^2}{6}\left(2\left[\chi_{,\mu},\left[\chi,\chi^{,\mu}\right]\right]+\left[\chi,\left[\chi,\Box\chi\right]\right]\right),
    \end{split}
\end{equation}
while by variation with respect to the transverse modes we obtain 
\begin{equation}
    \begin{split}
         \left(\Box+m^2\right)A_k^T&=P_{ki}^T\left\{2ig\left(-m^2\chi\chi_{,i}+m^2\left[A_i^T,\frac{1}{\Delta}\left(\Ddot{\chi}\right)\right]+2m^2\left[\dot{A}_i^T,\frac{1}{\Delta}\left(\dot{\chi}\right)\right]\right.\right.\\\\
         &\left.\left.+2\left[A_j^T,A_{i,j}^T\right]+\left[A_{j,i}^T,A_j^T\right]\right)-\frac{g^2m^2}{3}\left[\chi,\left[\chi_{,i},\chi\right]\right]\right\}.
    \end{split}
\end{equation}
In order to develop the perturbation theory, we will expand the longitudinal and transverse modes in powers of the coupling constant: 
\begin{equation}
    \chi=\chi^{(0)}+\chi^{(1)}+...\qquad\text{and}\qquad A_i^T=A_i^{T(0)}+A_i^{T(1)}+...
\end{equation}
where $\chi^{(0)}$ and $A_i^{T(0)}$ satisfy 
\begin{equation}
     \left(\Box+m^2\right)\chi^{(0)}=0\qquad\text{and}\qquad \left(\Box+m^2\right)A_i^{T(0)}=0.
\end{equation}
The most dominant contributions to the first-order corrections are given by 
\begin{equation}
    \left(\Box+m^2\right)\chi^{(1)}\sim2ig\left[A_i^T,\chi_{,i}\right]\sim\frac{g}{\left(mL\right)L^2}
\end{equation}
for the longitudinal modes. Hence, we can estimate
\begin{equation}
    \chi^{(1)}\sim\frac{g}{mL}.
\end{equation}
Since $\chi^{(0)}\sim\frac{1}{mL}$, we can see that these corrections are smaller than the kinetic term provided that $g\gg 1$. The first-order corrections to the transverse modes are given by 
\begin{equation}
    \begin{split}
         \left(\Box+m^2\right)A_k^{T(1)}\sim2igP_{ki}^T\left\{-m^2\chi^{(0)}\chi^{(0)}_{,i}+2\left[A_j^{T(0)},A_{i,j}^{T(0)}\right]+\left[A_{j,i}^{T(0)},A_j^{T(0)}\right]\right\}\sim\frac{g}{L^3}.
    \end{split}
\end{equation}
Therefore, we can estimate them as
\begin{equation}
    A_k^{T(1)}\sim\frac{g}{L}.
\end{equation}
This term is always smaller than the linear term. The second-order corrections to the longitudinal modes are given by 
\begin{equation}
    \left(\Box+m^2\right)\chi^{(2)}\sim\frac{g^2}{3}\left[\chi^{(0)}_{,\mu},\left[\chi^{(0)},\chi^{(0),\mu}\right]\right]\sim\frac{g^2}{\left(mL\right)^3L^2}.
\end{equation}
Therefore we can estimate them as 
\begin{equation}
    \chi^{(2)}\sim\frac{g^2}{\left(mL\right)^3}.
\end{equation}
They become of the same order as the linear term at the scale
\begin{equation}
    L_{str}\sim\frac{g}{m},
\end{equation}
which agrees with the strong coupling scale obtained from the Lagrangian density. The second-order corrections to the transverse modes satisfy \begin{equation}
    \left(\Box+m^2\right)A_k^{T(2)}\sim-\frac{g^2m^2}{3}P_{ik}^T\left\{\left[\chi^{(0)},\left[\chi^{(0)}_{,i},\chi^{(0)}\right]\right]\right\}\sim\frac{g^2}{\left(mL\right)L^3},
\end{equation}
and hence can be estimated as 
\begin{equation}
    A_k^{T(2)}\sim\frac{g^2}{\left(mL\right)L}.
\end{equation}
They become of the same order as the linear term $A_i^{T(0)}$ at the scales
\begin{equation}
    L\sim\frac{g^2}{m}.
\end{equation}
However, this scale is smaller than one of the  longitudinal modes. As the perturbation theory for the longitudinal modes breaks down before this scale is reached, it can no longer be trusted. In order to find the true corrections for the transverse modes, we need to evaluate the theory beyond the strong coupling scale of the longitudinal modes.
\subsection{Beyond the strong coupling scale}

Let us now consider the theory at the Vainshtein scale $L_{str}$. The analysis of the higher order terms shows that on this scale there is an infinite number of terms of the form 
\begin{equation}
    \mathcal L_{int}\supset \sum_{n=2}^{\infty}\frac{g^nm^2}{L^2}\chi^{n+2}
\end{equation}
which are just as relevant as the term $n=2$, despite being suppressed for $L>L_{str}$. The reason for this is the minimal level of quantum fluctuations of the longitudinal mode at the strong coupling scale:
\begin{equation}
   \delta\chi_{L_{str}}\sim\frac{1}{g},
\end{equation}
which implies that the expansion of $\zeta$ is no longer valid. Therefore, to analyse the theory beyond the strong coupling scale, we should leave $\zeta$ intact and return to the original Lagrangian density (\ref{eq::nonpertLagrangian}).\footnote{Note that it is still possible to expand the $\frac{1}{D}$ operator}. As a first step in determining the corrections to the transverse modes, we will estimate the minimum level of quantum fluctuations of the longitudinal modes. For energies $k^2\gg m^2$, the corresponding kinetic term is that of the nonlinear sigma model, and is given by 
\begin{equation}\label{eq::nonlinsigmamodel}
    \mathcal{L}_0^{\chi}\sim\frac{m^2}{g^2}\tr\left(\zeta^{\dagger}_{,\mu}\zeta^{,\mu}\right)
\end{equation}
Instead of working with the SU(2) matrix $\zeta$ which we have parametrized as
\begin{equation}
    \zeta=e^{-ig\chi},
\end{equation}
we will now consider its matrix elements 
\begin{equation}
    \zeta=\begin{pmatrix}
     \zeta_2^* & \zeta_1 \\\\
     -\zeta_1^* & \zeta_2
\end{pmatrix}.
\end{equation}
Here, $\zeta_1$ and $\zeta_2$ are two complex fields which satisfy 
\begin{equation}\label{eq::circle}
    |\zeta_1|^2+|\zeta_2|^2=1.
\end{equation}
Furthermore, we will substitute
\begin{equation}\label{eq::subst}
    \zeta_1=\rho_1e^{ig\theta_1}\qquad\text{and}\qquad \zeta_2=\rho_2e^{ig\theta_2}.
\end{equation}
Inserting (\ref{eq::subst}) into (\ref{eq::circle}), we obtain 
\begin{equation}
    |\rho_1|^2+|\rho_2|^2=1.
\end{equation}
Substituting further
\begin{equation}
    \rho_1=\rho\cos(g\sigma)\qquad\text{and}\qquad\rho_2=\rho\sin(g\sigma),
\end{equation}
 (\ref{eq::circle}) becomes
\begin{equation}
    |\rho|^2=1.
\end{equation}
Setting $\rho=1$, we obtain
\begin{equation}
    \zeta_1=\cos(g\sigma)e^{ig\theta_1}\qquad\text{and}\qquad \zeta_2=\sin(g\sigma)e^{ig\theta_2}.
\end{equation}
In terms of the fields $\sigma$, $\zeta_1$ and $\zeta_2$, the Lagrangian density (\ref{eq::nonlinsigmamodel}) becomes
\begin{equation}\label{eq::ElLag}
    \mathcal{L}_0^{\chi}=\frac{1}{2}\left[4m^2\partial_{\mu}\sigma\partial^{\mu}\sigma+f^2(\sigma)\partial_{\mu}\theta_1\partial^{\mu}\theta_1+p^2(\sigma)\partial_{\mu}\theta_2\partial^{\mu}\theta_2\right],
\end{equation}
where 
\begin{equation}
    f^2(\sigma)=4m^2\cos^2\left(g\sigma\right)\qquad\text{and}\qquad p^2(\sigma)=4m^2\sin^2\left(g\sigma\right).
\end{equation}
Clearly, these fields are not canonically normalised. In terms of the canonically normalised variables 
\begin{equation}
    \sigma_n=2m\sigma,\qquad\qquad \partial_{\mu}\theta_{1n}=f(\sigma) \partial_{\mu}\theta_{1},\qquad\text{and}\qquad \partial_{\mu}\theta_{2n}=p(\sigma) \partial_{\mu}\theta_{2},
\end{equation}
the Lagrangian density is given by 
\begin{equation}
    \mathcal{L}_0^{\chi}=\frac{1}{2}\left[\partial_{\mu}\sigma_n\partial^{\mu}\sigma_n+\partial_{\mu}\theta_{1n}\partial^{\mu}\theta_{1n}+\partial_{\mu}\theta_{2n}\partial^{\mu}\theta_{2n}\right].
\end{equation}
The minimal level of quantum fluctuations of the normalised fields for scales $k^2\sim\frac{1}{L^2}\gg m^2$ is given by 
\begin{equation}\label{eq::quaFLN}
    \delta\sigma_{nL}\sim\frac{1}{L},\qquad\qquad\delta\theta_{n1L}\sim\frac{1}{L},\qquad\text{and}\qquad \delta\theta_{n2L}\sim\frac{1}{L}.
\end{equation}
Estimating $\sin(g\sigma)\sim\cos(g\sigma)\sim\mathcal{O}(1)$, we find the minimal level of quantum fluctuations of the original fields:
\begin{equation}
  \delta\sigma_L\sim\frac{1}{2g}\frac{k}{k_{str}},\qquad\qquad\delta\theta_{1L}\sim\frac{1}{2g}\frac{k}{k_{str}},\qquad\text{and}\qquad\delta\theta_{2L}\sim\frac{1}{2g}\frac{k}{k_{str}}.
\end{equation}
Since we are considering the elements of $\zeta$, it is natural that we also consider the matrix elements of the transverse modes. We can write them in the following way:
\begin{equation}\label{eq::Amatrix}
    A_i^T=\begin{pmatrix}
     -\frac{G_i}{2} & -\frac{W_i^+}{\sqrt{2}} \\\\
     -\frac{W_i^-}{\sqrt{2}} & \frac{G_i}{2} 
\end{pmatrix},
\end{equation}
where $G_i$ is the real and $W^{\pm}_i$ complex vector fields, which are transverse:
\begin{equation}
    G_{i,i}=0\qquad\qquad\text{and}\qquad\qquad W^{\pm}_{i,i}=0.
\end{equation}
Substituting (\ref{eq::Amatrix}) into $\mathcal{L}_0^T$, we obtain
\begin{equation}
    \mathcal{L}_0^T=\frac{1}{2}\left[\partial_{\mu}G_i\partial^{\mu}G_i-m^2G_iG_i\right]+\partial_{\mu}W^+_i\partial^{\mu}W^-_i-m^2W^+_iW^-_i
\end{equation}
The kinetic terms imply
\begin{equation}
    \delta G_L\sim\frac{1}{L}\qquad\text{and}\qquad \delta W_L^{\pm}\sim\frac{1}{L}.
\end{equation}
Let us now analyse the interacting terms. We will start with the most problematic interaction. The leading divergences for the transverse modes within perturbation theory are due to the following term:
\begin{equation}
    \mathcal{L}^{T\chi}_{int}\supset-\frac{2im^2}{g}\tr\left(A_i^T\zeta^{\dagger}\zeta_{,i}\right).
\end{equation}
Expressed in the form of matrix elements, it is given by
\begin{equation}
    \begin{split}
        -\frac{2im^2}{g}\tr\left(A_i^T\zeta^{\dagger}\zeta_{,i}\right)=&-2im^2\left\{\frac{\sigma_{,i}}{\sqrt{2}}\left[W_i^-e^{ig(\theta_1+\theta_2)}-W_i^+e^{-ig(\theta_1+\theta_2)}\right]\right.\\\\
        &\left.+ i\theta_{1,i}\left[G_i\cos^2(g\sigma)-\frac{\cos(g\sigma)\sin(g\sigma)}{\sqrt{2}}\left(W_i^-e^{ig(\theta_1+\theta_2)}+W_i^+e^{-ig(\theta_1+\theta_2)}\right)\right]\right.\\\\
         &\left.+ i\theta_{2,i}\left[G_i\sin^2(g\sigma)+\frac{\sin(g\sigma)\cos(g\sigma)}{\sqrt{2}}\left(W_i^-e^{ig(\theta_1+\theta_2)}+W_i^+e^{-ig(\theta_1+\theta_2)}\right)\right]\right\}.
    \end{split}
\end{equation}
Or, in terms of the normalized fields, we have
\begin{equation}
    \begin{split}
        -\frac{2im^2}{g}\tr\left(A_i^T\zeta^{\dagger}\zeta_{,i}\right)\sim-im&\left\{\frac{\sigma_{n,i}}{\sqrt{2}}[W_i^-h-W_i^+h^*]+iG_i\left[\theta_{1n,i}\cos\left(\frac{g\sigma_n}{2m}\right)+\theta_{2n,i}\sin\left(\frac{g\sigma_n}{2m}\right)\right]\right.\\\\
        &\left.+\frac{i}{\sqrt{2}}\left[\theta_{2n,i}\cos\left(\frac{g\sigma_n}{2m}\right)-\theta_{1n,i}\sin\left(\frac{g\sigma_n}{2m}\right)\right][W_i^-h+W_i^+h^*]\right\}
    \end{split}
\end{equation}
 with
\begin{equation}
    h\sim e^{ig(\theta_1+\theta_2)}.
\end{equation}
Estimating 
\begin{equation}
    \sin(g\sigma)\sim\mathcal{O}(1),\qquad\qquad\cos(g\sigma)\sim\mathcal{O}(1)\qquad\text{and}\qquad h\sim\mathcal{O}(1),
\end{equation}
and taking into account the minimal level of quantum fluctuations for the fields (\ref{eq::quaFLN}), we can evaluate this term as 
\begin{equation}\label{eq::probint}
    -\frac{2im^2}{g}\tr\left(A_i^T\zeta^{\dagger}\zeta_{,i}\right)\sim g\frac{L}{L_{str}}\frac{1}{L^4}.
\end{equation}
We can see that the divergences in mass obtained in the framework of perturbation theory have now disappeared. For scales $L<L_{str}$, the most problematic term is now suppressed by the coupling constant and the ratio of the scale $L$ with the scale of the strong coupling. Therefore, this term decreases as we approach smaller scales or higher energies. Furthermore, this term also decreases as we approach a smaller mass, as in this case $L_{str}$ increases.
We will now show that the remaining interactions between longitudinal and transverse modes give a smaller contribution in comparison to the interaction we have now considered. First of all, we can see that these interactions contain the operator $\frac{1}{D}$. Taking into account the minimal level of quantum fluctuations for the transverse modes, this operator can always be expanded with each term in higher powers of the coupling constant being smaller than the previous one. Therefore, we can estimate this operator as 
\begin{equation}
    \frac{1}{D}\sim\frac{1}{-\Delta}.
\end{equation}
Defining 
\begin{equation}
    \Omega_0=\zeta^{\dagger}\dot{\zeta},
\end{equation}
we can write these interactions as
\begin{equation}\label{eq::remaining}
\begin{split}
    \mathcal{L}^{T\chi}_{int}&\supset\tr\left\{\frac{2im^4}{g}\Omega_0\frac{1}{\Delta}\left[\dot{A}_i^T,\frac{1}{\Delta}\left(\Omega_{0,i}\right)\right]+2m^2\Omega_0\frac{1}{\Delta}\left[\dot{A}_i^T,A_i^T\right]-m^4\Omega_0\frac{1}{\Delta}\left[A_i^T,\left[A_i^T,\frac{1}{\Delta}\left(\Omega_0\right)\right]\right]\right\}.
    \end{split}
\end{equation}
First we must to estimate $\Omega_{0}$. Let
\begin{equation}
    \Omega_{0}=\begin{pmatrix}
     \psi_{0} & \Phi_{0} \\\\
     -\Phi_{0}^* & -\psi_{0} 
\end{pmatrix},
\end{equation}
with $\mu=0,1,2,3$. The components of this matrix are given by  
\begin{equation}
    \psi_{0}=-ig(\dot{\theta}_1\cos^2(g\sigma)+\dot{\theta}_2\sin^2(g\sigma))=-\frac{ig}{2m}(\dot{\theta}_{1n}\cos(g\sigma)+\dot{\theta}_{2n}\sin(g\sigma))\sim\frac{g}{mL^2},
\end{equation}
and
\begin{equation}
    \begin{split}
        \Phi_{\mu}=&g[-\dot{\sigma}+i(\dot{\theta}_1-\dot{\theta}_2)\cos(g\sigma)\sin(g\sigma)]e^{i\theta_1+\theta_2}\sim\frac{g}{mL^2}.
    \end{split}
\end{equation}
Therefore, we can estimate 
\begin{equation}
    \Omega_{0}\sim\frac{g}{mL^2}.
\end{equation}
However, we have to be careful, because the interactions (\ref{eq::remaining}) include also the derivatives of $\Omega_{0}$.
Each derivative acting on it adds a factor of $\frac{g}{mL^2}$. For example, we have
\begin{equation}
    \partial_{\mu}\Omega_0\sim\frac{g^2}{\left(mL\right)^2L^2}.
\end{equation}
 Having this in mind, we can evaluate the interactions (\ref{eq::remaining}) as 
 \begin{equation}
 \begin{split}
     &\tr\left\{\frac{2im^4}{g}\Omega_0\frac{1}{\Delta}\left[\dot{A}_i^T,\frac{1}{\Delta}\left(\Omega_{0,i}\right)\right]\right\}\sim g^3\left(\frac{L}{L_{str}}\right)^5,\qquad\qquad \tr\left\{2m^2\Omega_0\frac{1}{\Delta}\left[\dot{A}_i^T,A_i^T\right]\right\}\sim g^2\left(\frac{L}{L_{str}}\right)^2\\\\
    &\text{and}\qquad \tr\left\{-m^4\Omega_0\frac{1}{\Delta}\left[A_i^T,\left[A_i^T,\frac{1}{\Delta}\left(\Omega_0\right)\right]\right]\right\}\sim g^4\left(\frac{L}{L_{str}}\right)^6.
 \end{split}
 \end{equation}
 It should be noted that we have here taken into account that we are evaluating the corrections for the transverse modes. Hence, in this case, the estimates must be equivalent to the equations of motion. As a result, all derivatives and $\frac{1}{-\Delta}$ must act on $\Omega_0$. As we can see, these corrections are clearly subdominant compared to (\ref{eq::probint}). Therefore, the leading order corrections to the transverse modes due to the longitudinal modes are given by (\ref{eq::probint}). Equivalently, the corrections at the level of the equations of motion are given by 
 \begin{equation}
    A_i^{T(1)}\sim-i\frac{m^2}{g}\zeta^{\dagger}\zeta_{,i}\sim\frac{g}{L^3}\frac{L}{L_{str}}.
\end{equation}
Therefore, we have obtained the massless Yang-Mills theory up to small corrections that disappear in the massless limit.

\section{Conclusion}
It is well-established that both massless and massive Yang-Mills theories are unitary and renormalizable, if the mass of the gauge bosons is generated by the Brout-Englert-Higgs mechanism \cite{masslessYM, HiggsYM, Higgs1, Higgs2, EB }. If, on the other hand, a mass term is added \textit{by hand}, the \textit{standard perturbative approach} suggests that the theory is neither renormalizable nor unitary, since the perturbative series is singular in the massless limit. In other words, there is a discontinuity when mass is set to zero. This can be observed already from the propagator of the vector fields, that tends to a constant at high energies and becomes infinite in the case of vanishing mass. Although in \cite{Veltman1970} it was proposed that the perturbative series might be re-summed, this proposal was soon discarded in \cite{vDVZ} because of the existence of a discontinuity in the 1-loop corrections to the propagator.
As a consequence, the massless limit appeared not to be smooth.
This is due to the longitudinal mode, a degree of freedom which is absent in the massless theory. 
\\
Nevertheless, the purpose of this paper was to show that these issues are merely an artefact of the standard perturbative approach. To begin with, we have pursued a conjecture given in \cite{Vainshtein}, namely, a smooth massless limit could be possible outside of the perturbation theory. We showed that, due to the nonlinear terms, the properly defined longitudinal modes become strongly coupled at the scale coinciding with the unitarity violation scale. As a result, singularities in mass that have emerged in the corrections to the transverse mode can no longer be trusted - the perturbation theory for the longitudinal modes breaks down before these singularities would become even relevant at all.  
\\
It is interesting to compare this theory to the one that in addition  contains also a Higgs field. If we were to rewrite the latter theory in terms of the gauge invariant variables \cite{Cosmo}, and set the Higgs field to a constant, both theories would coincide. However, with a non-constant field, the two theories differ because of the non-linear structure of the Lagrangian, resulting from a constraint (\ref{eq::circle}). Whereas this constraint is absent in the presence of the Higgs field, in the theory with a mass term added \textit{by hand} it remains, and hence the resulting Lagrangian for the longitudinal modes corresponds to a nonlinear sigma model. 
\\
At the Vainshtein scale, this model cannot be perturbatively expanded. Nonetheless, using the minimal level of quantum fluctuations of the fields, we found that it is still possible to develop the perturbation theory for transverse modes beyond the strong coupling scale. In fact, their corrections which were previously singular in the mass are now suppressed by the strong coupling scale. When the mass approaches zero, the Vainshtein scale approaches infinity and these corrections disappear. Thus, at high energies, the longitudinal modes completely decouple from the transverse modes, which remain in the weakly coupled regime. Therefore, the massless theory is restored up to small corrections. This leads us to the conclusion that the conjecture made in \cite{Vainshtein} was correct. The massless limit of the massive Yang-Mills theory is smooth.

\section*{Acknowledgements}
 \textit{ I would like to thank Ali H. Chamseddine and Tobias B. Russ for their contributions to the early work on this topic, Stefan Hofmann, Thomas Steingasser and Jan-Niklas Toelstede for useful discussions, and Andrea Quadri for helpful correspondence. In particular, I am thankful to Arkady Vainshtein for illuminating discussions. Finally, I am especially grateful to my PhD supervisor Slava Mukhanov for guidance and discussions while working on this thesis. \\This work was partially supported by the Deutsche Forschungsgemeinschaft
(DFG, German Research Foundation) under Germany’s Excellence Strategy – EXC-2111 – 390814868.}

\newpage

\section*{\textbf{A} $\quad$ The full Lagrangian density up to $\mathcal{O}\left(g^2\right)$ and the Feynman rules}\label{section:A}
Here we will present the full Lagrangian density up to $\mathcal{O}\left(g^2\right)$ for linearly defined transverse and longitudinal modes, without the $k^2\gg m^2$ approximation. We will also present the corresponding Feynman rules. The following shorthand notation will be used:
\begin{equation}
    D[\chi]\equiv\frac{-\Delta}{-\Delta+m^2}\left(\chi\right)\quad\quad\text{and}\quad\quad F[\chi]\equiv\frac{\Delta+m^2}{-\Delta+m^2}\chi,
\end{equation}
where  $\frac{1}{-\Delta+m^2}$ should be understood in the momentum space. Then, the Lagrangian density up to $\mathcal{O}(g^2)$ is given by
\begin{equation}
    \mathcal{L}= \mathcal{L}_0+ \mathcal{L}_{g}+ \mathcal{L}_{g^2},\qquad \text{where}
\end{equation}
\begin{equation*}
    \begin{split}
        &\mathcal{L}_0=\mathcal{L}_{0\chi}+\mathcal{L}_{0A^T}\\\\
        &\mathcal{L}_{0\chi}=-\frac{m^2}{2}\chi^a(\Box+m^2)D[\chi^a]\quad\quad\quad
        \mathcal{L}_{0A^T}=-\frac{1}{2}A_i^{Ta}(\Box+m^2)A_i^{Ta}\\\\
        &\mathcal{L}_g=\mathcal{L}_{g\chi}+\mathcal{L}_{gA^T}+\mathcal{L}_{g2\chi A^T}+\mathcal{L}_{g\chi 2A^T}\\\\
        &\mathcal{L}_{g\chi}=gm^2\varepsilon^{abc}D[\dot{\chi}^a]\frac{1}{-\Delta+m^2}\left(\dot{\chi}^b_{,i}\right)\chi^c_{,i}\quad\quad\quad
        \mathcal{L}_{gA^T}=g\varepsilon^{abc}A_{j,i}^{Ta}A_i^{Tb}A_j^{Tc}\\\\
        &\mathcal{L}_{g2\chi A^T}=g\varepsilon^{abc}\left\{D[\dot{\chi}^a]\left[  \frac{m^2}{-\Delta+m^2}\left(\dot{\chi}^b_{,i}\right)A_i^{Tc}+\dot{A}_i^{Tb}\chi^c_{,i}\right] +A_{j,i}^{Ta}\chi^b_{,i}\chi^c_{,j} \right\} \\\\
        &\mathcal{L}_{g\chi2A^T}=g\varepsilon^{abc}\left(D[\dot{\chi}^a]\dot{A}_i^{Tb}A_i^{Tc}+A_{j,i}^{Ta}\chi^b_{,i}A_j^{Tc}\right)\\\\
        &\mathcal{L}_g^2=\mathcal{L}_{g^2\chi}+\mathcal{L}_{g^2A^T}+\mathcal{L}_{g^23\chi A^T}+\mathcal{L}_{g^22\chi2A^T}+\mathcal{L}_{g^2\chi3A^T}\\\\
        &\mathcal{L}_{g^2\chi}=-\frac{1}{2}g^2\varepsilon^{fab}\varepsilon^{fcd}\left\{\chi_{,i}^bF[\dot{\chi}^a_{,i}]-\Delta(\chi^b)D[\dot{\chi}^a]\right\}\frac{1}{-\Delta+m^2}\left\{\chi_{,j}^dF[\dot{\chi}^c_{,j}]-\Delta(\chi^d)D[\dot{\chi}^c]\right\}\\
        &\quad\quad\quad+\frac{1}{2}g^2\varepsilon^{fab}\varepsilon^{fcd}\left[D[\dot{\chi}^a]\chi_{,i}^bD[\dot{\chi}^c]\chi_{,i}^d-\frac{1}{2}\chi_{,i}^a\chi_{,j}^b\chi_{,i}^c\chi_{,j}^d\right]\\\\
        &\mathcal{L}_{g^2A^T}=-\frac{1}{2}g^2\varepsilon^{fab}\varepsilon^{fcd}\left\{\left[\dot{A}_i^{Ta}A^{Tb}_i\right]\frac{1}{-\Delta+m^2}\left[\dot{A}_j^{Tc}A^{Td}_j\right]+\frac{1}{2}A_i^{Ta}A_j^{Tb}A_i^{Tc}A_j^{Td}\right\}\\\\
        &\mathcal{L}_{g^23\chi A^T}=-g^2\varepsilon^{fab}\varepsilon^{fcd}\left\{\dot{A}_i^{Ta}\chi^b_{,i}+A_i^{Tb}F[\dot{\chi}_{,i}^a]\right\}\frac{1}{-\Delta+m^2}\left\{\chi_{,j}^dF[\dot{\chi}_{,j}^c]-\Delta(\chi^d)D[\dot{\chi}^c]\right\}\\
        &\quad\quad\quad\quad\quad+g^2\varepsilon^{fab}\varepsilon^{fcd}\left\{D[\dot{\chi}^a]A_i^{Tb}D[\dot{\chi}^c]\chi_{,i}^d-\chi_{,j}^aA_i^{Tb}\chi_{,i}^d\chi_{,j}^c\right\}\\\\
        &\mathcal{L}_{g^22\chi2A^T}=-\frac{1}{2}g^2\varepsilon^{fab}\varepsilon^{fcd}\left\{2\left[\dot{A}_i^{Ta}A_i^{Tb}\right]\frac{1}{-\Delta+m^2}\left[\dot{A}_j^{Tc}\chi_{,j}^d+A_j^{Td}F[\dot{\chi}^c_{,j}]\right]\right.\\
        &\quad\quad\quad\quad\left.+\left[\dot{A}_i^{Ta}\chi_{,i}^b+A_i^{Tb}F[\dot{\chi}^a_{,i}]\right]\frac{1}{-\Delta+m^2}\left[\dot{A}_j^{Tc}\chi_{,j}^d+A_j^{Td}F[\dot{\chi}^c_{,j}]\right]\right\}\\
        &\quad\quad\quad\quad-\frac{1}{2}g^2\varepsilon^{fab}\varepsilon^{fcd}\left\{-D[\dot{\chi}^a]A_i^{Tb}A_i^{Td}D[\dot{\chi}^c]+A_i^{Ta}A_j^{Tb}\chi^c_{,i}\chi^d_{,j}+A_i^{Ta}A_i^{Tc}\chi^b_{,j}\chi^d_{,j}+A_i^{Ta}A_j^{Td}\chi^b_{,j}\chi^c_{,i}\right\}
    \end{split}
\end{equation*}

\begin{equation*}
    \begin{split}
        &\mathcal{L}_{g^2\chi3A^T}=-g^2\varepsilon^{fab}\varepsilon^{fcd}\left\{\left[\dot{A}_i^{Ta}A_i^{Tb}\right]\frac{1}{-\Delta+m^2}\left[\dot{A}_j^{Tc}\chi_{,j}^d+A_j^{Td}F[\dot{\chi}^c_{,j}]\right]+A_i^{Ta}A_j^{Tb}A_i^{Tc}\chi^d_{,j}\right\}.
    \end{split}
\end{equation*}
Let us now consider the Feynman rules. We can see that the longitudinal modes are not normalised, and normalisation according to (\ref{eq::cannorm}) would put the kinetic term into the form of a standard massive scalar field. However, substituting the normalised variables in the interacting part of the Lagrangian would complicate the expressions for the vertices. Therefore, we will work with the original longitudinal mode and express the Feynman rules through it. It is easy to verify that the resulting expressions for the diagrams are independent of this choice. To define the asymptotic states and find the expression for the propagator of the original longitudinal mode, one simply performs a canonical normalisation and then rewrites the expressions in terms of the original mode. In the following, we will only present the Feynman rules up to and including $\mathcal{O}(g)$.
The propagators for the longitudinal and transverse modes are each given in momentum space by \newline 
\vspace{-5mm}
\begin{wrapfigure}[3]{z}{0.4\textwidth}
    \centering
    \includegraphics[width=0.15\textwidth]{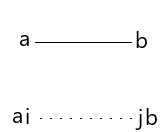}
\end{wrapfigure}
\begin{equation*}
    \begin{split}
        &\Delta_{\chi}^{ab}(k)=\frac{|\vec{k}|^2+m^2}{m^2|\vec{k}|^2}\frac{i\delta^{ab}}{k^2-m^2}\\\\
        &\Delta_{ij}^{Tab}(k)=\left(\delta_{ij}-\frac{k_ik_j}{|\vec{k}|^2}\right)\frac{i\delta^{ab}}{k^2-m^2}
    \end{split}
\end{equation*}
At $\mathcal{O}(g)$ we have the following vertices, with all momenta outgoing which have been ordered according to the number of the longitudinal lines. 
\begin{figure}[h]
\includegraphics[width=16cm]{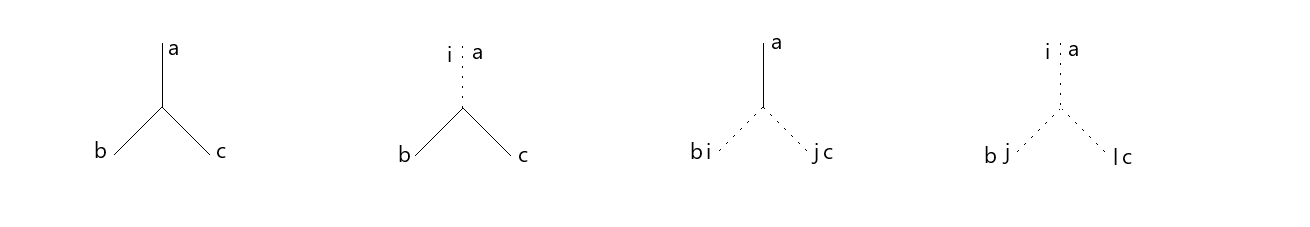}
\centering
\end{figure}
\hfill\break
\vspace{-10mm}
\begin{equation}
    \begin{split}
        V^{abc}_{3\chi}(k,p,q)=&igm^2\varepsilon^{abc}\left[\frac{|\vec{k}|^2k_0p_iq_i}{|\vec{k}|^2+m^2}\left(\frac{p_0}{|\vec{p}|^2+m^2}-\frac{q_0}{|\vec{q}|^2+m^2}\right)+ \frac{|\vec{p}|^2p_0k_iq_i}{|\vec{p}|^2+m^2}\left(\frac{q_0}{|\vec{q}|^2+m^2}-\frac{k_0}{|\vec{k}|^2+m^2}\right)+\right.\\
      & \left.+\frac{|\vec{q}|^2q_0k_ip_i}{|\vec{q}|^2+m^2}\left(\frac{k_0}{|\vec{k}|^2+m^2}-\frac{p_0}{|\vec{p}|^2+m^2}\right)\right]\\\\
      V^{abc}_{i,2\chi}(k_{(i)},p,q)=&-g\varepsilon^{abc}\left[
     k_{\mu}\left(q^{\mu}p_i-p^{\mu}q_i\right)+m^2k_0\left(\frac{p_0q_i}{|\vec{p}|^2+m^2}-\frac{q_0p_i}{|\vec{q}|^2+m^2}\right)
   +\frac{m^2p_0q_0\left(|\vec{p}|^2q_i-|\vec{q}|^2p_i\right)}{(|\vec{p}|^2+m^2)(|\vec{q}|^2+m^2)}\right]\\\\
    V_{ij,\chi}^{abc}(k,p_{(i)},q_{(j)})=&-ig\varepsilon^{abc}\delta_{ij}\left[\frac{|\vec{k}|^2k_0}{|\vec{k}|^2+m^2}(p_0-q_0)-k_l(p-q)_l\right]\\\\
     V_{ijl}^{abc}(k,p,q)=&g\varepsilon^{abc}\left[\delta_{il}(k-q)_j+\delta_{ij}(p-k)_l+\delta_{jl}(q-p)_i\right]
    \end{split}
\end{equation}
\section*{\textbf{B} $\quad$ Massless Yang-Mills theory}\label{section:B}
In the massive Yang-Mills theory, we will only work with longitudinal and transverse modes. When calculating the corrections to the propagator of the transverse modes, we want to compare these contributions with the massless theory. We can see that the propagator of the transverse modes in the massive case corresponds to the propagator of the Coulomb gauge in the massless limit. The massless theory is evaluated along these lines in the radiation gauge, where only the propagating degrees of freedom are taken into account. 
The action is given by 
\begin{equation}
    S=-\frac{1}{2}\int d^4x\tr(F_{\mu\nu}F^{\mu\nu}).
\end{equation}
First, we will decompose the vector field in terms of temporal and spatial part, $A_0^a$ and $A_i^a$ with $a=1,2,3$. Then, we obtain
\begin{equation}
\begin{split}
    S&=\int d^4x\left\{\frac{1}{2}\left[A_0^a(-\Delta)A_0^a+2A_0^a\Dot{A}_{i,i}^{a}\right]
    +\frac{1}{2}(\dot{A}_i^a\dot{A}_i^a+A_{j,i}^aA_{i,j}^a-A_{j,i}^aA_{j,i}^a)\right.\\
    &\left.+g\epsilon^{abc}(-A_0^bA_i^c\dot{A}_i^a-A_i^bA_0^cA_{0,i}^a+A_i^bA_j^cA_{j,i}^a)
    -\frac{g^2}{4}\epsilon^{abc}\epsilon^{aed}(-2A_0^bA_i^cA_0^eA_i^d+A_i^bA_j^cA_i^eA_j^d)\right\}.
\end{split}
\end{equation}
As a next step, we can impose the radiation gauge condition:
\begin{equation*}
    A_{i,i}^a=0.
\end{equation*}
As in the massive case, the temporal part does not propagate because no time derivatives act on it. Therefore, the theory has only two propagation modes. Let us integrate out the non-propagating component like in the massive case. The component $A_0$ satisfies a system of constraints (one for each $a=1,2,3$):
\begin{equation}
\begin{split}
    -\Delta A_0^a=-\dot{A}_{i,i}^a-g\epsilon^{abc}\dot{A}_i^bA_i^c-g\epsilon^{abc}A_i^bA_{0,i}^c-g\epsilon^{abc}\partial_i(A_i^bA_0^c)-g^2\epsilon^{fbc}\epsilon^{fad}A_0^bA_i^cA_i^d
\end{split}
\end{equation}
Up to $\mathcal{O}\left(g\right)$, the solution of the constraints can be can be evaluated as 
\begin{equation}
\begin{split}
    A_0^a&= -g\varepsilon^{abc}\frac{1}{-\Delta}\left[\dot{A}_i^bA_i^c\right]
    +2g^2\varepsilon^{fab}\varepsilon^{fcd}\frac{1}{-\Delta}\left[(A_i^b\partial_i)\frac{1}{-\Delta}\left(\dot{A}_j^cA_j^d\right)\right]
    +\mathcal{O}(g^3),
\end{split}
\end{equation}
and the Lagrangian density is given by 
\begin{equation}
\begin{split}
        &\mathcal{L}=\mathcal{L}_0+\mathcal{L}_{int},\qquad\text{where}\\\\
        &\mathcal{L}_0=-\frac{1}{2}A_i^{Ta}(\Box)A_i^{Ta}\qquad\text{and}\\\\
        &\mathcal{L}_{int}=g\varepsilon^{abc}A_{j,i}^{Ta}A_i^{Tb}A_j^{Tc}-\frac{1}{2}g^2\varepsilon^{fab}\varepsilon^{fcd}\left\{\left[\dot{A}_i^{Ta}A^{Tb}_i\right]\frac{1}{-\Delta}\left[\dot{A}_j^{Tc}A^{Td}_j\right]+\frac{1}{2}A_i^{Ta}A_j^{Tb}A_i^{Tc}A_j^{Td}\right\}.
\end{split}
\end{equation}
We can see from the kinetic term  that the minimal level of quantum fluctuations of the transverse modes is given by
\begin{equation}
    \delta A_L^{Ta}\sim\frac{1}{L},  \quad\text{}\quad a=1,2,3.
\end{equation}
Also, we can notice that the vertex at $\mathcal{O}(g)$ exactly agrees with the vertex $V_{ijl}^{abc}(k,p,q)$ of the massive theory, while the propagator corresponds to the one of transverse modes in the massive case with mass set to zero:
\begin{equation*}
    \Delta_{ij}^{Tab}(k)=\left(\delta_{ij}-\frac{k_ik_j}{|\vec{k}|^2}\right)\frac{i\delta^{ab}}{k^2}.
\end{equation*}
\section*{\textbf{C}$\quad$ Can we estimate $L^T_{str}$ from the Feynman diagrams?}
We have seen that for a linear decomposition of the vector field, the transverse modes enter the strong coupling regime at scales $L^T_{str}$. However, the literature has so far had no mention of unitarity breakdown at the same scales. Rather, it has been shown that the unitarity breakdown occurs at energies $k_u\sim\frac{m}{g}$ \cite{VeltmanReiff}. This corresponds to the two-loop correction of the propagator of transverse modes, while at one loop there are no inverse powers of mass, but the amplitude does not coincide with that of the corresponding massless theory \cite{vDVZ}. 
With this in mind, one might wonder whether the Feynman rules we obtained for the theory with linearly defined transverse and longitudinal modes would lead to different results. The strong coupling was found at the level of the equations of motion for the transverse modes. Therefore, it is reasonable to suspect that the violation of unitarity corresponding to the same energies could occur as a result of the corrections to the propagator of the transverse modes. First, we will analyse the one-loop contribution to the transverse modes, and show that the results agree with \cite{vDVZ}. After confirming that there is no violation of unitarity for one-loop corrections, we will outline the proof of the absence of $L^T_{str}$ for two loops.
\\\\
In a first step, we will show that the imaginary part of the corrections to the propagator of the transverse modes in the massless limit does not agree with the massless theory. In this search, we will follow an approach similar to \cite{Veltman1968, VeltmanReiff, vDVZ} and calculate the imaginary part of the corrections at the propagator of the vector field using the cutting rules \cite{Cutkosky}. The reason why we focus on the imaginary part of the diagrams is its close connection to the unitarity of the S-matrix. The condition that the S-matrix is unitary means that 
\begin{equation}
    S^{\dag}S=1
\end{equation}
is satisfied, where $S$ denotes the S-matrix. We can express the S-matrix through the $T$-matrix by
\begin{equation}
    S=1+iT.
\end{equation}
Then, if $\ket{a},\ket{b}, \ket{c}$ denote the states of the system, the unitarity condition can be expressed as  
\begin{equation}
    2\text{Im}\bra{b}T\ket{a}=-\sum_c\bra{b}T\ket{c}\bra{c}T^{\dag}\ket{a}.
\end{equation}
From the perspective of diagrams, this corresponds to a cutting equation. Therefore, this statement is equivalent to saying that the imaginary part of the diagram is given by the sum over all possible cuts of the diagram \cite{Diagrammar}. This is just another way of expressing the Optical theorem \cite{Peskin}.
Let us now analyse the corrections to the propagator of transverse modes at one loop in massive and massless case. The Feynman diagrams for the massive theory are given in \ref{section:A}. The massless theory will be evaluated in the Coulomb gauge, with the Feynman rules given in \ref{section:B}. 
In the massive case there are five diagrams contributing to the corrections to the transverse modes:
\begin{figure}[H]
\includegraphics[width=14cm]{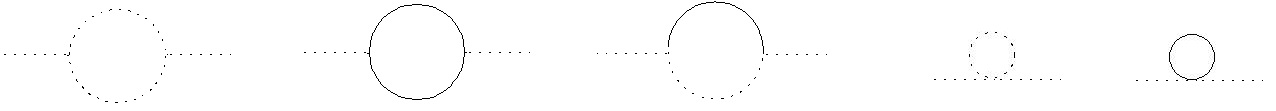}
\centering
\end{figure}
\vspace{-4mm}
\hfill\break Here, the dotted line represents the propagator for the transverse modes, while the full line corresponds to the longitudinal modes. 
In the massless case only the first diagram is present. The imaginary part of the diagrams corresponds to the sum over all possible cuts \cite{Diagrammar}. Hence, the imaginary part of the last two diagrams automatically vanishes, and we are left with the following cuts:
\begin{figure}[h]
\includegraphics[width=9cm]{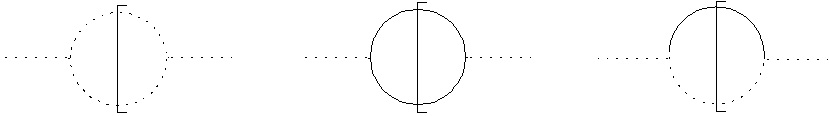}
\centering
\end{figure}
\vspace{-4mm}
\hfill\break
Cutting a certain propagator line corresponds to a replacement of the propagator with 
\begin{equation*}
    \begin{split}
        &\tilde{\Delta}_{\chi}^{ab}(k)=2\pi\theta(k_0)\delta(k^2-m^2)\frac{|\vec{k}|^2+m^2}{m^2|\vec{k}|^2},\\\\
        &\tilde{\Delta}^{Tab}_{ij}(k)=2\pi\theta(k_0)\delta(k^2-m^2)\delta^{ab}\left(\delta_{ij}-\frac{k_ik_j}{|\vec{k}|^2}\right),
    \end{split}
\end{equation*}
in the case of longitudinal and transverse modes respectively, 
and for a positive energy flow \cite{Diagrammar}. In other words, cutting a propagator line sets the momenta corresponding to it on-shell. This is the generalisation of the Cutkosky rules for our case. 
In the massless case one has to make a following replacement
\begin{equation*}
    \tilde{\Delta}^{Tab}_{ij}(k)=2\pi\theta(k_0)\delta(k^2)\delta^{ab}\left(\delta_{ij}-\frac{k_ik_j}{|\vec{k}|^2}\right).
\end{equation*}
In all of the diagrams, we will denote the external momenta with $p$. These are already on-shell, meaning that $p_i\varepsilon^a_{i}(p)=0$, where the $\varepsilon^a_{i\sigma}(p)$ are polarisation vectors appearing due to the transverse modes. The polarisation vectors satisfy
\begin{equation*}
    \sum_{\sigma=1,2}\varepsilon^a_{i\sigma}(p)\varepsilon^b_{j\sigma}(p)=\left(\delta_{ij}-\frac{p_ip_j}{|\vec{p}|^2}\right)\delta^{ab}.
\end{equation*}
First we will analyse the diagram containing only transverse modes. The imaginary part is given by
\begin{equation}
\begin{split}
    &\text{Im}\Gamma^T=\int\frac{d^4k}{(2\pi)^3}\int\frac{d^4q}{(2\pi)^3}(2\pi)^4\delta^{(4)}(p-k-q)\theta(k_0)\delta(k^2-m^2)\theta(q_0)\delta(q^2-m^2)\varepsilon^a_{i}(p)\varepsilon_{j}^b(-p)T^{ab}_{ij},\\
    &\text{where}\qquad T^{ab}_{ij}=V^{acd}_{ikl}(-p,k,q)V^{bcd}_{jnz}(p,-k,-q)\left(\delta_{kn}-\frac{k_kk_n}{|\vec{k}|^2}\right)\left(\delta_{lz}-\frac{q_lq_z}{|\vec{q}|^2}\right),
\end{split}
\end{equation}
for the massive case, and 
\begin{equation}
    \text{Im}\Gamma^T=\int\frac{d^4k}{(2\pi)^3}\int\frac{d^4q}{(2\pi)^3}(2\pi)^4\delta^{(4)}(p-k-q)\theta(k_0)\delta(k^2)\theta(q_0)\delta(q^2)\varepsilon^a_{i}(p)\varepsilon_{j}^b(-p)T^{ab}_{ij},
\end{equation}
for the massless fields, with $T^{ab}_{ij}$ given in the expression for the massive fields. 
We can see that the vertex, which contains only transverse modes, coincides in both the massive and massless cases. As expected, there are no mass divergences in these expressions because there are no longitudinal modes.  Consequently, the first diagram exactly cancels with the corresponding massless diagram for $m\to0$. 
\newline
The last diagram only provides terms that are multiplied by powers of the mass and thus, it will disappear in the massless limit. This can be easily seen from the vertices of the diagram. Let $g$ be the internal moments corresponding to the longitudinal propagator and $k$ correspond to the transverse ones. These are put on-shell, as the cut passes through their propagator lines, i.e. $k^2=q^2=m^2$. Therefore, one of the vertices takes the form
\begin{equation}
    V_{il}^{cad}=-igm^2\varepsilon^{cad}\delta_{il}\frac{1}{|\vec{q}|^2+m^2}\left(2q_nk_n+m^2|\vec{q}|^2\right).
\end{equation}
This cancels with the $\frac{1}{m^2}$ term arising from the longitudinal modes. However, since we have two vertices, the second one will give an additional $m^2$, and thus the contribution will vanish in the massless limit. 
\newline
In contrast to the first and last diagrams, the second one will give a difference when compared to the massless theory.
It's imaginary part is given by 
\begin{equation}
    \begin{split}
        \text{Im}\Gamma_L=\int\frac{d^4k}{(2\pi)^3}\int\frac{d^4q}{(2\pi)^3}&(2\pi)^4\delta^{(4)}(p-k-q)\theta(k_0)\delta(k^2-m^2)\theta(q_0)\delta(q^2-m^2)\varepsilon^a_{i}(p)\varepsilon_{j}^b(-p)I^{ab}_{ij},
    \end{split}
\end{equation}
where 
\begin{equation}
    I^{ab}_{ij}=-\frac{1}{4}\delta^{ce}\delta^{df}\frac{|\vec{k}|^2+m^2}{|\vec{k}|^2}\frac{|\vec{q}|^2+m^2}{|\vec{q}|^2}V^{acd}_{i,2\chi}(-p,k,q)V^{bef}_{j,2\chi}(p,-k,-q).
\end{equation}
 For $k^2=q^2=m^2$ we have 
\begin{equation}
    \begin{split}
        V^{acd}_{i,2\chi}(-p,k,q)=-gm^2\varepsilon^{acd}k_i\left[1+\frac{2m^2k_0q_0}{(|\vec{k}|^2+m^2)(|\vec{q}|^2+m^2)}\right],
    \end{split}
\end{equation}
and similar for the other vertex. Keeping only the relevant terms, we obtain
\begin{equation}
    I^{ab}_{ij}=\frac{1}{2}g^2\delta^{ab}k_ik_j.
\end{equation}
This contribution, which is purely due to the longitudinal modes, is clearly preserved in the massless limit. With this result, we are in agreement with the \cite{vDVZ}. Moreover, we can observe that the inverse powers of the mass are not present, and therefore there is no possibility of the unitarity violation. We can note that the vertices that constitute the second diagram correspond to the term in the Lagrangian which was responsible for the strong coupling scale. Therefore, we will now examine whether there is a possibility that the unitarity scale arises at the level of the two loops, where the diagrams are formed from this vertex. 
\\\\
At two loops, there are two diagrams that could give us the unitarity violation corresponding to the energy scales of the strong coupling. These are
\begin{figure}[H]
\includegraphics[width=9cm]{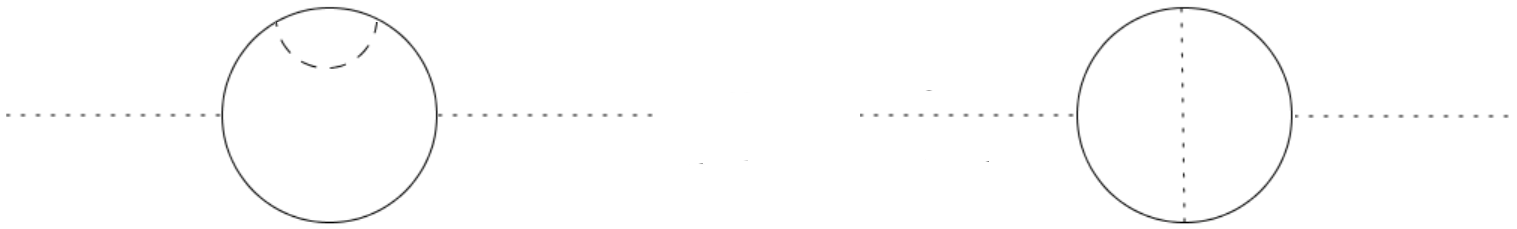}
\centering
\end{figure}
The cuts of the first diagram are given by 
\begin{figure}[H]
\includegraphics[width=12cm]{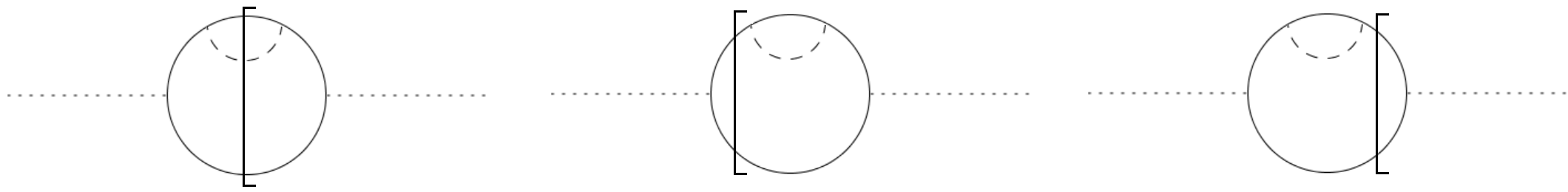}
\centering
\end{figure}
while for the second diagram we have 
\begin{figure}[H]
\includegraphics[width=16cm]{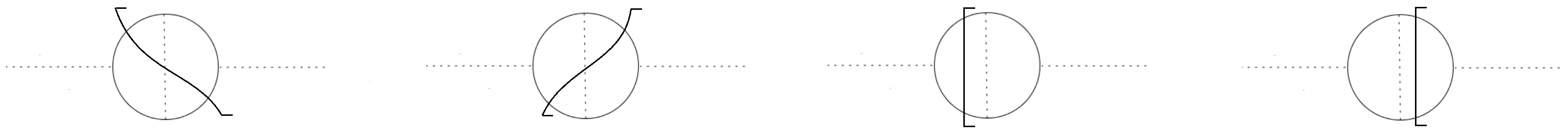}
\centering
\end{figure}
The largest possible mass divergence for these diagrams is $\sim\frac{g^4}{m^8}$, which means that there is a possibility for the existence of  $L^T_{str}$. However, it is easy to show that this is not the case. To show it, we need to confirm that at least one of the vertices $V_{i,2\chi}^{abc}(k_{(i)},p,q)$ will be proportional to $m^2$ at the lowest order. Let us consider the first diagram. Let $p$ be the external momenta, and $k$ and $l$ the internal momenta over which we integrate, with $k$ on the lowest line and $l$ on the line above. We can first do a trick by inserting 
\begin{equation}\label{eq::trick}
    \int d^4k\int d^4l=\int d^4k\int d^4l\int d^4q \delta^{(4)}\left(p-k-l-q\right)
\end{equation}
into the expression for the imaginary part of the diagram. Cutting a line means that the corresponding momenta are set  on-shell. In all three cases with cuts we therefore have $k^2=m^2$. Let us now look at the first vertex appearing in the first two diagrams. Neglecting the terms proportional in to $m^2$, it is given by
\begin{equation}
    V_{i,2\chi}^{abc}(-p_{(i)},k,p-k)\sim g\varepsilon^{abc}\left(-p_{\mu}k^{\mu}p_i-p^2k_i\right)=-g\varepsilon^{abc}m^2k_i,
\end{equation}
where we have used that the external momentum is on-shell, meaning that $\varepsilon_i^a(p)p_i=0$ and $p^2=m^2$. 
An analogous calculation applies to the last diagram, except that it is now more convenient to look at the last vertex rather than the first. Therefore, the first diagram cannot provide the scale of the unitarity scale which would match the strong coupling scale. Let us now look at the second diagram, and perform the same trick (\ref{eq::trick}) to analyse the cuts. The last two cuts cannot give $L^T_{str}$. This is due to the first (or last) vertex, next to which a cut is being made. This sets all three lines around the vertex on-shell, which makes the vertex contain only the terms with $m^2$. The first two cuts for the second diagram are again equivalent. Let $l$ lie on the line which passes through the loop. The cut over it leaves us with the transverse projector in momentum space, for which we have 
\begin{equation}
    l_i\left(\delta_{ij}-\frac{l_il_j}{|\vec{l}|^2}\right)=0.
\end{equation}
Let us now consider the vertex at the bottom of this line. If $k$ is the loop momentum exiting the vertex, we have 
\begin{equation}
     V_{i,2\chi}^{abc}(l_{(i)},k,-l-k)\sim g\varepsilon^{abc}l_{\mu}\left(l^{\mu}k_i-k^{\mu}l_i\right).
\end{equation}
The last term vanishes since it is contracted with the transverse projector. Since the cut is made through the middle line, this sets $l$ on shell, and hence this vertex does not have terms which are multiplied by mass. The analysis of the two loop corrections for the transverse modes therefore suggests that $L^T_{str}$  does not appear in the diagrams. 
\\\\
Note that the diagrams presented here refer to $A^T$ and $\chi$ components, with the $A_0$ component integrated out. To rewrite the results in terms of the $A_{\mu}$ fields, one can consult the following chapter. 

\section*{\textbf{D}$\quad$ Conversion of diagrams}
Here we will present the rules of transforming a manifestly covariant diagram into a set of manifestly non-covariant ones, rewritten only in terms of the longitudinal and transverse modes (We will refer to the latter one as $A^T\chi$ formulation) For simplicity, here we will ignore the colour indices. 
\newline
\begin{wrapfigure}[13]{z}{0.3\textwidth} 
    \centering
    \includegraphics[width=0.2\textwidth]{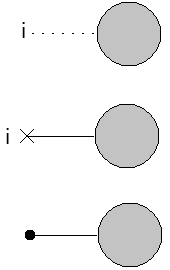}
\end{wrapfigure}
In order to do it, we first have to take into account that the external lines of manifestly covariant diagram carry external indices $\mu=0,1,2,3$.
 Based on this, there are 3 relevant types of external propagators in $A^T\chi$ formulation
\begin{enumerate}
    \item transverse 
    \item longitudinal with a cross, which comes with a factor $-ip_i$
    \item longitudinal with a circle, which comes with a factor $-ip_0A(p)$
\end{enumerate}
where $A(p)=\frac{\vec{p}^2}{\vec{p}^2+m^2}$ and big grey circle corresponds to the rest of the diagram. 
These factors should be multiplied with the corresponding type of the propagator
\begin{equation}
    \begin{split}
        &\Delta_{ij}^T(p)=(\delta_{ij}-\frac{k_ik_j}{\Vec{k}^2})\frac{i}{k^2-m^2} \quad\quad\text{for transverse modes}\\
        &\Delta_{\chi}=\frac{1}{m^2A(k)}\frac{i}{k^2-m^2}\quad\quad\text{for longitudinal modes}
    \end{split}
\end{equation}
For the longitudinal modes there are two kinds, since they come from $A_i$ and $A_0$.
Now, in an manifestly covariant theory, the diagram can be decomposed into transverse and longitudinal parts according to the following rules
\begin{itemize}
    \item Draw all combinations of the external propagators.
    \item Connect them with all possible vertices. 
    \item The value of a given diagram then follows the usual Feynman rules, the only difference being that a diagram should now be multiplied by factors from three types of propagators. 
\end{itemize}

Let's consider as an example 1-loop diagram. Then all possible diagrams one could form are given by 

\begin{figure}[h]
\includegraphics[width=16cm]{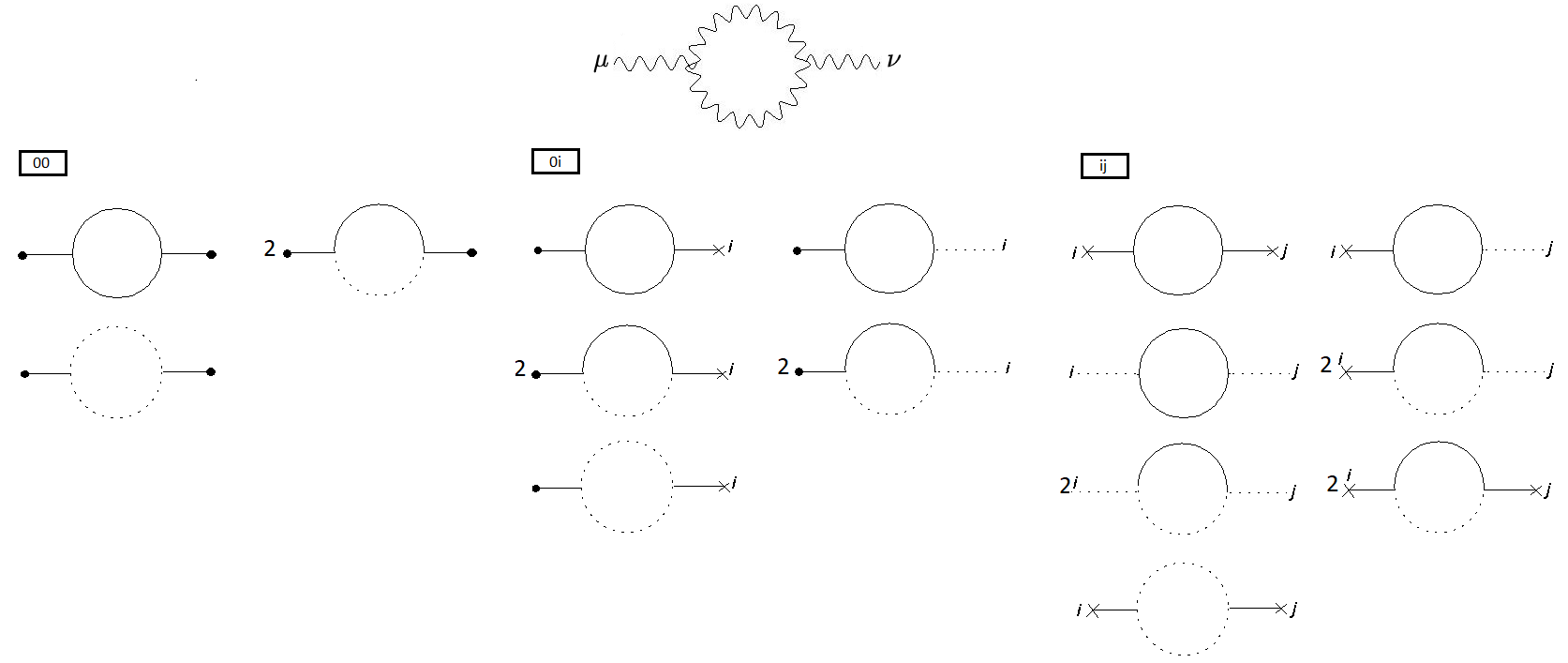}
\centering
\end{figure}



\newpage
\begin{thebibliography}{9}

\bibitem{Glashow}
S. L. Glashow, \textit{Partial Symmetries of Weak Interactions,} Nucl. Phys. \textbf{22} (1961), 579-588.

\bibitem{Delburgo}
R. Delbourgo and S. Twisk and G. Thompson, \textit{MASSIVE YANG-MILLS THEORY: RENORMALIZABILITY VERSUS UNITARITY}, Int. J. Mod. Phys. A \textbf{3} (1988), 435

\bibitem{Proca}
A. Proca, \textit{Sur la theorie ondulatoire des electrons positifs et negatifs}, J. Phys. Radium \textbf{7} (1936), 347-353.

\bibitem{BoulGil}
 D. G. Boulware and W. Gilbert, \textit{Connection between Gauge Invariance and Mass,} Phys. Rev. \textbf{126} (1962), 1563-1567.

\bibitem{Stueckelberg}
E. C. G. Stueckelberg, \textit{Die Wechselwirkungskräfte in der Elektrodynamik und in der Feldtheorie der Kräfte,} Helv. Phys. Acta 11, (1938), 225.

\bibitem{Kunimasa}
T. Kunimasa and T.Goto, \textit{Generalization of the Stueckelberg Formalism to the Massive Yang-Mills Field,} Prog. Theor. Phys. \textbf{37} (1967), 452-464.

\bibitem{SalamKomar}
A. Komar and A. Salam, \textit{Renormalization problem for vector meson theories,} Nucl. Phys. \textbf{21} (1960), 624-630.

\bibitem{UmezawaKamefuchi} 
H. Umezawa and S. Kamefuchi,  \textit{Equivalence theorems and renormalization problem in vector field theory
(The Yang–Mills field with non-vanishing masses)} Nucl. Phys. \textbf{23} (1961), 399-429.

\bibitem{Ionides}
P.A. Ionides, \textit{General equivalence theorem in vector field theory}, Nucl. Phys. \textbf{28} (1961), 662-664.

\bibitem{Salam}
A. Salam, \textit{Renormalizability of Gauge Theories}, Phys. Rev. \textbf{127} (1962), 331-334.

\bibitem{Boulware}
 D. G. Boulware,\textit{Renormalizeability of massive non-Abelian gauge fields: A functional integral approach} Annals Phys. \textbf{56} (1970), 140-171.
 
 \bibitem{Veltman1970}
M.J.G. Veltman, \textit{Generalized ward identities and yang-mills fields}, Nucl. Phys. B \textbf{21} (1970), 288-302.

\bibitem{Gegelia}
J. Gegelia and G. Japaridze, \textit{On renormalizability of the effective field theory of massive Yang-Mills fields,} Mod. Phys. Lett. A \textbf{27} (2012), 1250128 [arXiv:1109.3880 [hep-th]].


\bibitem{cuttingrules}
M.J.G. Veltman, \textit{Unitarity and causality in a renormalizable field theory with unstable particles}, Physica \textbf{29} (1963), 186-207.

 \bibitem{Cutkosky}
 R.E. Cutkosky, \textit{Singularities and discontinuities of Feynman amplitudes}, J. Math. Phys. \textbf{1} (1960), 429-433.
 
 
\bibitem{Veltman1968}
M.J.G. Veltman, \textit{Perturbation theory of massive Yang-Mills fields}, Nucl. Phys. B \textbf{7} (1968), 637-650.

\bibitem{VeltmanReiff}
J. Reiff, M.J.G. Veltman, \textit{Massive Yang-Mills fields}, Nucl. Phys. B \textbf{13} (1969), 545-564.

\bibitem{Bell}
J.S. Bell, \textit{HIGH-ENERGY BEHAVIOUR OF TREE DIAGRAMS IN GAUGE THEORIES}, Nucl. Phys. B \textbf{60} (1973), 427-436.

\bibitem{Smith}
C.H. Llewellyn Smith, \textit{High-Energy Behavior and Gauge Symmetry}, Phys. Lett. B \textbf{46} (1973), 233-236.

\bibitem{Cornwall}
 J.M. Cornwall, D.N. Levin and G. Tiktopoulos, \textit{Derivation of Gauge Invariance from High-Energy Unitarity Bounds on the S Matrix,} Phys. Rev. D \textbf{10} (1974), 1145
[erratum: Phys. Rev. D \textbf{11} (1975), 972].

\bibitem{Vainshtein}
A. I. Vainshtein and I. B. Khriplovich, \textit{On the zero-mass limit and renormalizability in the theory of massive yang-mills field,} Yad. Fiz. \textbf{13} (1971), 198-211. 


\bibitem{vDVZ}
H. van Dam and M. J. G. Veltman, \textit{Massive and massless Yang-Mills and gravitational fields,} Nucl. Phys. B \textbf{22} (1970), 397-411.


\bibitem{Wong}
S.Wong, \textit{Massless limit of the massive Yang-Mills field,} Phys. Rev. D \textbf{3} (1971), 945-952
[erratum: Phys. Rev. D \textbf{3} (1971), 3243-3243].
 
\bibitem{SlavFad}
A. A. Slavnov and L.D. Faddeev, \textit{Massless and massive Yang-Mills fields,} Theor. Math. Phys. \textbf{3} (1970), 312-316.


\bibitem{BassSch}
E. Schr\"{o}dinger and L. Bass, \textit{Must the photon mass be zero?}, Proc. Roy. Soc. (London) A232, (1955), 1.


\bibitem{FierszPauli}
M. Fierz and W. Pauli, \textit{On relativistic wave equations for particles of arbitrary spin in an electromagnetic field,}  Proc. Roy. Soc. Lond. A \textbf{173} (1939), 211-232.

\bibitem{Zakharov}
V.I. Zakharov, \textit{Linearized gravitation theory and the graviton mass} JETP Lett. \textbf{12} (1970), 312.

\bibitem{Iwasaki}
Y. Iwasaki, \textit{Consistency condition for propagators,} Phys. Rev. D \textbf{2} (1970), 2255-2256.


\bibitem{VainshteinMeh}
A. I. Vainshtein, \textit{To the problem of nonvanishing gravitation mass,} Phys. Lett. B \textbf{39} (1972), 393-394.



\bibitem{PertNep}
C. Deffayet, G. R. Dvali, G. Gabadadze and A. I. Vainshtein, \textit{Nonperturbative continuity in
graviton mass versus perturbative discontinuity,} Phys. Rev. D \textbf{65} (2002), 044026 [arXiv:hep-th/0106001].


\bibitem{Mimetic}
A. H. Chamseddine and V. Mukhanov, \textit{Mimetic Massive Gravity: Beyond Linear Approximation,} JHEP \textbf{06} (2018), 062 [arXiv:1805.06598 [hep-th]].

\bibitem{Gruzinov}
A. Gruzinov, \textit{On the graviton mass,} New Astron. \textbf{10} (2005), 311-314 [arXiv:astro-ph/0112246 [astro-ph]].

\bibitem{Slavnov1972}
A. A. Slavnov, \textit{Massive gauge fields,} Teor. Mat. Fiz. \textbf{10} (1972), 305-328.

 
\bibitem{QuadriUnit}
R. Ferrari and A. Quadri, \textit{Physical unitarity for massive non-Abelian gauge theories in the Landau gauge: Stueckelberg and Higgs,} JHEP \textbf{11} (2004), 019 [arXiv:hep-th/0408168 [hep-th]].




\bibitem{Ferrari}
D. Bettinelli, R. Ferrari and A. Quadri, \textit{A Massive Yang-Mills Theory based on the Nonlinearly Realized Gauge Group,} Phys. Rev. D \textbf{77} (2008), 045021 [arXiv:0705.2339 [hep-th]].

\bibitem{QuadriOneloop}
D. Bettinelli, R. Ferrari and A. Quadri, \textit{One-loop self-energy and counterterms in a massive Yang-Mills theory based on the nonlinearly realized gauge group,} Phys. Rev. D \textbf{77} (2008), 105012
[erratum: Phys. Rev. D \textbf{85} (2012), 129901]
[arXiv:0709.0644 [hep-th]].

\bibitem{Quad}
Private correspondence with A. Quadri. 

\bibitem{Dragon}
N. Dragon, T. Hurth and P. van Nieuwenhuizen,
\textit{Polynomial form of the Stuckelberg model,}
Nucl. Phys. B Proc. Suppl. \textbf{56} (1997), 318-321.

\bibitem{Ruegg}
H. Ruegg and M. Ruiz-Altaba, \textit{The Stueckelberg field,} Nucl. Phys. B Proc. Suppl. \textbf{56} (1997), 318-321 
[arXiv:hep-th/9703017 [hep-th]].


 
\bibitem{FradkinTy}
E. S. Fradkin, and I. V. Tyutin, \textit{Feynman rules for the massless Yang–Mills field renormalizability of the theory of the massive Yang–Mills field}, Phys. Lett. B \textbf{30} (1969), 562-563.


\bibitem{QFTCS}
 V. Mukhanov, S. Winitzki, \textit{Introduction to Quantum Effects in Gravity}, Cambridge University Press (2007).

 
 \bibitem{masslessYM}
G.'t Hooft, \textit{Renormalization of Massless Yang-Mills Fields,} Nucl. Phys. B \textbf{33} (1971), 173-199. 

\bibitem{HiggsYM}
G.'t Hooft,\textit{Renormalizable Lagrangians for Massive Yang-Mills Fields,}
Nucl. Phys. B \textbf{35} (1971), 167-188.

\bibitem{Higgs1}
P. W. Higgs, \textit{Broken symmetries, massless particles and gauge fields,} Phys. Lett. \textbf{12} (1964), 132-133.

\bibitem{Higgs2}
P. W. Higgs, \textit{Broken Symmetries and the Masses of Gauge Bosons,} Phys. Rev. Lett. \textbf{13} (1964), 508-509.

\bibitem{EB}
F. Englert and R. Brout, \textit{Broken Symmetry and the Mass of Gauge Vector Mesons,} Phys. Rev. Lett. \textbf{13} (1964), 321-323.

\bibitem{Cosmo}
 V. Mukhanov, \textit{Physical Foundations of Cosmology}, Cambridge University Press (2005).
 
\bibitem{Diagrammar}
 M. Veltman, \textit{Diagrammatica: The Path to Feynman Diagrams (Cambridge Lecture Notes in Physics)}, Cambridge: Cambridge University Press, (1994).
 
\bibitem{Peskin}
M. E. Peskin and D. V. Schroeder, \textit{An Introduction to quantum field theory}, Addison-Wesley (1995).



\end{thebibliography}
\end{document}